%% file: main.tex
\pdfoutput=1
\documentclass[journal]{IEEEtran}
\IEEEoverridecommandlockouts

\usepackage{cite}
\usepackage{amsmath, amssymb, amsfonts}
\usepackage{algorithmic}
\usepackage{graphicx}
\usepackage{textcomp}
\usepackage{xcolor}

\usepackage{siunitx}
\usepackage{subcaption}
\usepackage{commath}
\usepackage{tabularx, booktabs}
\usepackage{bbding}
\usepackage{pifont}
\usepackage{float}
\DeclareMathOperator*{\argmax}{argmax}

\newcommand{\RH}{\mathrm{RH}} 
\newcommand{\PU}{\mathrm{PU}}
\usepackage{makecell} 
\newcommand\given[1][]{\:#1\vert\:}
\begin{document}
\title{Optimizing Throughput Performance in \\ Distributed MIMO Wi-Fi Networks using \\ Deep Reinforcement Learning}

\author{Neelakantan Nurani~Krishnan,~\IEEEmembership{Student Member,~IEEE,}
        Eric~Torkildson,~\IEEEmembership{Member,~IEEE,}
        Narayan~B.~Mandayam,~\IEEEmembership{Fellow,~IEEE,}
        Dipankar~Raychaudhuri,~\IEEEmembership{Fellow,~IEEE,}
        Enrico~Rantala,~\IEEEmembership{Member,~IEEE,}
        and~Klaus~Doppler,~\IEEEmembership{Senior Member,~IEEE}
\thanks{N. Nurani Krishnan, N. B. Mandayam, and D. Raychaudhuri are with the Wireless Information Network Laboratory (WINLAB), Department of Electrical and Computer Engineering, Rutgers University, North Brunswick, NJ, USA. E-mail: \{neel45, narayan, ray\}@winlab.rutgers.edu} %
\thanks{E. Torkildson, E. Rantala, and K. Doppler are with Nokia Bell Labs, Sunnyvale, CA, USA. E-mail: \{eric.torkildson, enrico-henrik.rantala, klaus.doppler\}@nokia-bell-labs.com}
\thanks{This work has been submitted to the IEEE for possible publication. Copyright may be transferred without notice, after which this version may no longer be accessible.}} %

\maketitle

\begin{abstract}
This paper explores the feasibility of leveraging concepts from deep reinforcement learning (DRL) to enable dynamic resource management in Wi-Fi networks implementing distributed multi-user MIMO (D-MIMO). D-MIMO is a technique by which a set of wireless access points are synchronized and grouped together to jointly serve multiple users simultaneously. This paper addresses two dynamic resource management problems pertaining to D-MIMO Wi-Fi networks: (i) channel assignment of D-MIMO groups, and (ii) deciding how to cluster access points to form D-MIMO groups, in order to maximize user throughput performance. These problems are known to be NP-Hard and only heuristic solutions exist in literature. We construct a DRL framework through which a learning agent interacts with a D-MIMO Wi-Fi network, learns about the network environment, and is successful in converging to policies which address the aforementioned problems. Through extensive simulations and on-line training based on D-MIMO Wi-Fi networks, this paper demonstrates the efficacy of DRL in achieving an improvement of 20\% in user throughput performance compared to heuristic solutions, particularly when network conditions are dynamic. This work also showcases the effectiveness of DRL in meeting multiple network objectives simultaneously, for instance, maximizing throughput of users as well as fairness of throughput among them. 
\end{abstract}

\input{introduction.tex}

\input{overview_of_dmimo_wifi.tex}
\input{motivating_scenarios.tex}
\input{drl_framework}

\input{results.tex}
\input{discussion.tex}
\input{conclusion.tex}

\bibliographystyle{IEEEtran} 
\bibliography{main.bib}

\end{document}

%% file: introduction.tex
\section{Introduction} 
\label{sec:introduction}
Emerging data-intensive applications such as Augmented/Virtual Reality (AR/VR) and 8K video will drive the throughput requirements of next-generation Wi-Fi networks. To meet rising throughput demands over time, Wi-Fi has steadily added support for wider channel bandwidths, including \SI{40}{\MHz} in 802.11n, \SI{80/160}{\MHz} in 802.11ac, and \SI{320}{\MHz} now under consideration \cite{xtreme_throughput}. Another approach to achieving higher throughput is that of reducing the distance between neighboring Wi-Fi access points (APs). This allows each AP to serve a smaller area and provide its users with higher average SNR. In practice, however, these two approaches of wider channels and denser networks are at odds. Interference between closely-spaced networks limits availability of the $80$--\SI{160}{\MHz} channels, which are accessed on a best-effort basis, causing devices fall back to narrower $40$ or \SI{20}{\MHz} channels. 

Distributed MIMO (D-MIMO), also known as Network MIMO, is a potential solution to these issues as studied in \cite{balan2012achieving, neel2018dmimoo, lopez2019ieee} and the references within. A D-MIMO system consists of several time and phase-synchronized APs that jointly transmit and receive signals, thereby acting as a single spatially-distributed virtual antenna array to simultaneously serve multiple users (hence the name distributed multi-user MIMO). The cooperation between APs reduces intra-network interference and improves spatial reuse of wireless channels. Distributed MIMO as a technology and the merits of its use have been understood and studied extensively in literature through both simulations and implementations. Our earlier work \cite{neel2018dmimoo} proposed the use of distributed MIMO in the context of dense Wi-Fi networks, along with the necessary architectural changes, and demonstrated a $2.5\times$ improvement in the throughput performance of users with D-MIMO compared to baseline deployments. 

The objective of this paper is to add intelligence to D-MIMO Wi-Fi networks to empower them with an autonomous adaptation to dynamic network scenarios. This work addresses two important dynamic resource management problems in D-MIMO networks, which are listed below, by leveraging principles from deep reinforcement learning:
\begin{enumerate}
    \item \textbf{\emph{Channel assignment problem}}: If there are $N$ D-MIMO groups and $K$ available channels ($K<N$), what is the best channel assignment policy to maximize user throughput performance? 
    \item \textbf{\emph{AP clustering problem}}: How should APs be clustered together to form D-MIMO groups to best serve users in the network? How should this grouping be updated in response to changing user distributions? 
\end{enumerate}
\subsection{Related Work} 
The first mentioned problem is essentially the typical channel assignment problem in any Wi-Fi network, which is known to be \textbf{\emph{NP-Hard}}, and many heuristic solutions exist in literature \cite{chieochan2010channel}. On the other hand, there exists limited literature on clustering of APs for distributed MIMO. Authors of \cite{zhang2013nemox, zhang2009networked, neel2018dmimoo} grouped nearby antennas/APs into one cluster but the grouping was static. Another strategy to group antennas was on a per-packet basis \cite{papadogiannis2008dynamic, liu2009improved, weber2011self} which needed full channel state information (CSI) from the users. Although the clustering was dynamic, the proposed solution was difficult to implement in practice because of the time varying nature of CSI. References \cite{shi2000normalized, lin2017user} showed that the problem of determining the best AP clustering policy was \textbf{\emph{NP-Complete}} (that is, both NP and NP-Hard), and the authors in \cite{lin2017user} performed dynamic clustering of APs based on user uplink traffic distribution. The authors, however, did not consider downlink service to users and variability in user distribution. Since both the aforementioned problems are known to be NP-Hard and they get exacerbated when the network environment is dynamic, it is worthwhile to investigate if learning-based methods can address these problems wherein an agent learns about the dynamics of the network environment and acts accordingly. 

Deep learning and reinforcement learning have emerged in recent years as technologies of valuable importance in a gamut of fields, be it image or pattern recognition, robotics, or even DNA synthesis. Deep reinforcement learning (DRL) is a synergistic combination of these two techniques in which deep neural networks are used in the training process to improve the learning speed and performance of reinforcement learning algorithms, especially in high-dimensional state and action spaces. DRL has been used to play a variety of games and the agents have been successful in outperforming human players and achieving superhuman scores \cite{mnih2015human}. They have proved their effectiveness in wireless communications as well. Some examples of using DRL in communications include proactive data caching and computation overloading in mobile edge networks \cite{zhong2018deep, he2018integrated}, preservation of wireless connectivity \cite{wang2017autonomous}, traffic engineering and resource scheduling \cite{xu2018experience, zhao2018deep}, and enabling multiple access in wireless networks \cite{yu2018deep}. Reference \cite{luong2018applications} comprehensively reviews the existing applications of DRL in communications and networking found in literature.

\subsection{Summary of Contributions}
The main contributions of this paper include building model-free DRL agents to address the aforementioned dynamic resource management problems in D-MIMO Wi-Fi networks. Specifically, the agents perform \emph{on-line training}
\begin{enumerate}
    \item to determine the policy of channel assignment for D-MIMO groups that maximizes user throughput performance, even in the presence of time-varying external Wi-Fi interference, 
    \item to meet multiple network objectives simultaneously (for instance, maximize throughput of users as well as fairness of throughput distribution among them), and  
    \item to determine the AP clustering policy that maximizes user throughput performance in response to non-uniform spatial distribution of users as well as user mobility. 
\end{enumerate}

The emphasis on the term on-line training highlights the fact that the agent learns about the network environment in real-time, that is, with every learning episode. To the best of our knowledge, this is the first attempt at using DRL in the context of D-MIMO Wi-Fi networks, particularly for meeting multiple objectives and in the problem of AP clustering. We also consider practical network scenarios like non-uniform spatial distribution of users and user mobility. This work demonstrates that DRL agents with fairly simple implementations, in terms of number of hidden layers and the implemented learning algorithm, attain an improvement of $\sim20\%$ in user throughput performance in D-MIMO Wi-Fi networks compared to existing heuristic solutions.

\subsection{Organization of the paper}
Section~\ref{sec:overview_of_dmimo_wifi} reviews D-MIMO Wi-Fi and describes how the D-MIMO architecture helps improve throughput performance in dense Wi-Fi networks. Section~\ref{sec:motivating_scenarios} provides some example scenarios to motivate the use of DRL in D-MIMO Wi-Fi networks. Section~\ref{sec:drl_framework} gives an introduction to DRL and its associated terminology, a walk-through of an example learning episode, and the reasoning behind the choice of learning agents to address the current problems of interest. Section~\ref{sec:results} describes the simulation and learning setup, and provides a detailed account of the results obtained from on-line training based on simulations of various network scenarios. Section~\ref{sec:discussion} provides an account of the key takeaways from the results, the lessons learned from implementing the DRL framework, and future research directions in the realm of DRL aided wireless networks. Section~\ref{sec:conclusion} concludes the paper.

%% file: overview_of_dmimo_wifi.tex
\section{Overview of D-MIMO Wi-Fi} 
\label{sec:overview_of_dmimo_wifi}

This section briefly reviews the architecture of D-MIMO Wi-Fi, and discusses the motivation for implementing a D-MIMO Wi-Fi network. 

\subsection{Architecture Overview}
\label{architecture}

\begin{figure} 
\centering
\includegraphics[width=0.81\linewidth]{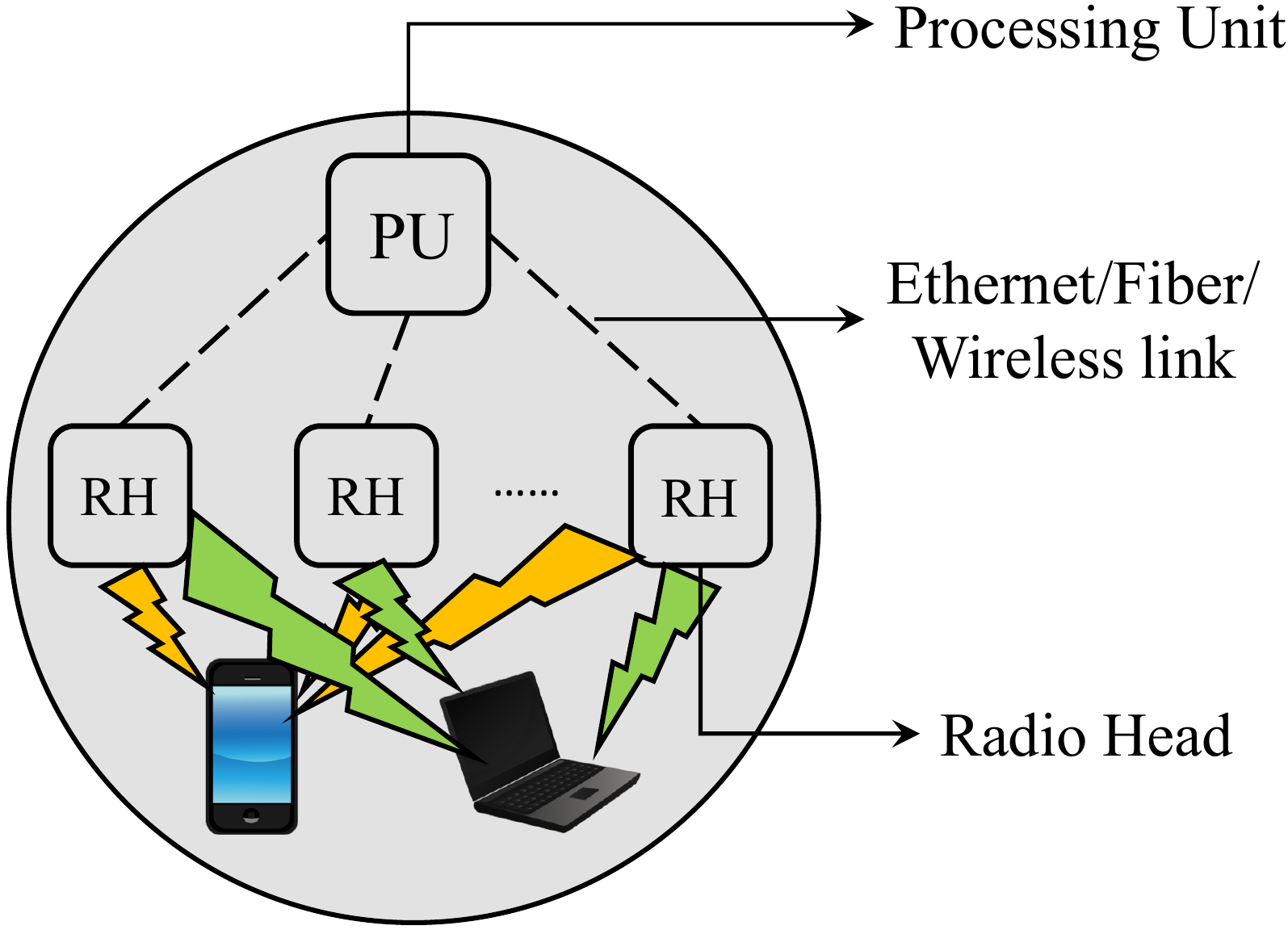}
\caption{Representative architecture of a D-MIMO group. A group with M $\RH$s, with N antennas per $\RH$, can support M$\times$N simultaneous downlink streams.}
\label{fig:dmimo_architecture}
\end{figure}

\begin{figure*}[t] 
	\begin{subfigure}[t]{.47\linewidth}
		\centering
		\includegraphics[width=0.75\linewidth]{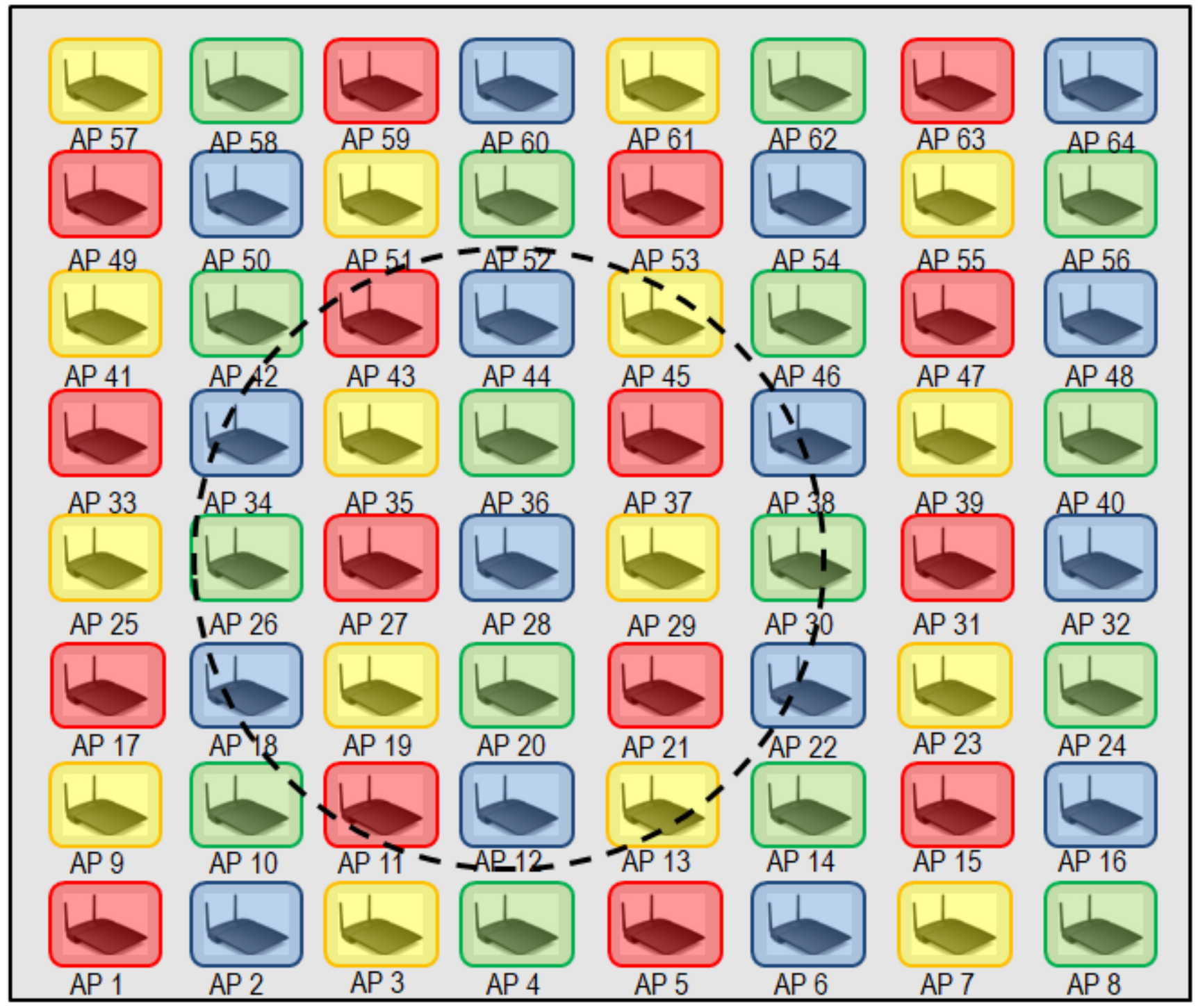}
		\caption{Baseline configuration -- Dense deployment of Wi-Fi APs with each AP assigned one channel}
		\label{fig:baseline_topology}
	\end{subfigure} \hfill
	\begin{subfigure}[t]{.47\linewidth}
		\centering
		\includegraphics[width=0.75\linewidth]{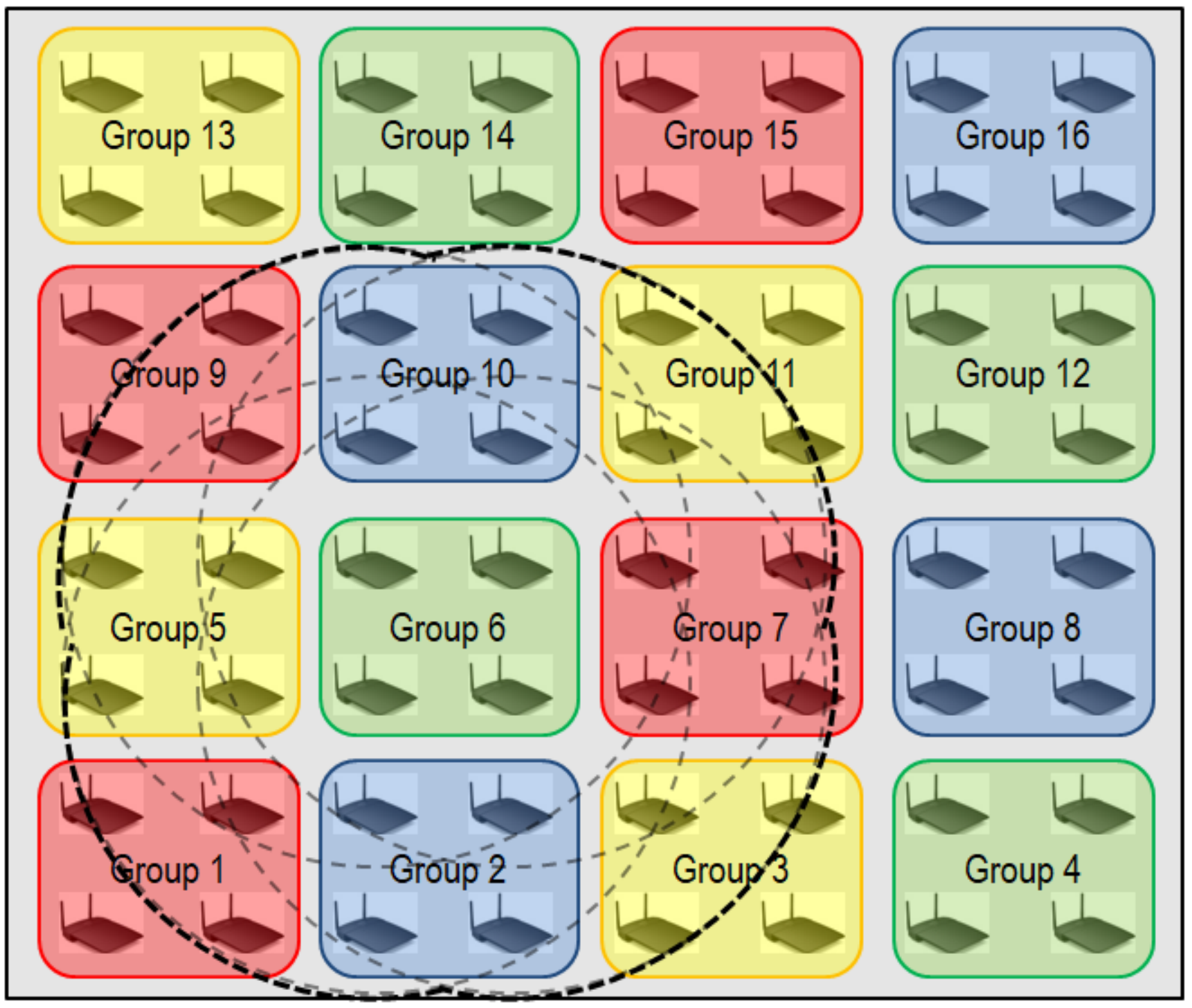}
		\caption{D-MIMO configuration -- $\RH$s are divided into groups and available channels are distributed among the groups}
		\label{fig:dmimo_topology}
	\end{subfigure} \hfill
	
	\caption{Cartoon example of dense deployment of Wi-Fi APs and D-MIMO $\RH$s. Frequency reuse factor is four with each channel represented by a unique color. Dotted circle represents the hearing range of an AP/$\RH$. }
\end{figure*}

The basic idea of D-MIMO is to divide the functionality of a Wi-Fi AP into two entities, radiohead ($\RH$) and processing unit ($\PU$) (see Fig.~\ref{fig:dmimo_architecture}). A $\RH$ is an external radio front end unit with one or more antennas providing a baseband signal interface for the $\PU$. The $\PU$ encompasses all the functionalities of an AP that are not in the $\RH$. The $\PU$ maintains time and phase synchronization in $\RH$s which may be connected to the $\PU$ using wired/wireless links.

\subsection{Motivation}
The trend in wireless network design recently has been dense access point deployments to increase system capacity. Keeping this in mind, the following two aspects motivate D-MIMO implementations. 

\begin{itemize}
\item \textbf{Wi-Fi densification}:

802.11ac/ax networks operate in the unlicensed 5 GHz frequency range, which is divided into a limited number of channels typically shared by multiple APs. As the inter-AP distance is reduced to improve SNR, more APs will share the same channel, resulting in less channel accesses per AP, and more control and management frames in total. Observe from Fig.~\ref{fig:baseline_topology} that \emph{hearing range} of an AP (the distance at which its transmissions reach above the clear channel assessment threshold) includes multiple other APs and their associated users in the same channel. This results in reduced transmission opportunities for all these APs. Deploying D-MIMO (see Fig.~\ref{fig:dmimo_topology}), however, employs coordination among APs and results in fewer neighboring networks on the same channel while still preserving the desired SNR. 

\item \textbf{Improved MIMO channel conditioning}: 

As described in \cite{dai2007some}, co-located antenna systems can suffer from low channel rank and correlated large scale fading experienced by all the transmit antennas. On the other hand, due to separation of transmitters in D-MIMO, spatial correlation of channels is reduced and hence the channel matrices have better conditioning. D-MIMO systems also achieve macro-diversity protection from all links having similar deep large scale fading. 

\end{itemize}

Implementing D-MIMO in the context of Wi-Fi networks invites us to rethink some fundamental Wi-Fi concepts, including but not limited to the carrier sense multiple access protocol with collision avoidance which mediates channel access among Wi-Fi access points, and MU-MIMO user selection in the downlink without incurring high channel sounding overhead. Standards compatible light-weight solutions to the aforementioned challenges were provided in our previous work \cite{neel2018dmimoo}. The proposed solutions as well as the general architectural change allowed D-MIMO to achieve a gain of $2.5\times$ in median throughput performance of users in comparison to a baseline dense deployment of APs as well as an improvement of $191\%$ in average user throughput. 

%% file: motivating_scenarios.tex
\section{Use of Reinforcement Learning in D-MIMO Wi-Fi Networks}
\label{sec:motivating_scenarios}
The scenario described in \cite{neel2018dmimoo} was a fairly simple deployment of a D-MIMO Wi-Fi network in which the users were static and distributed uniformly, and the network was not subject to any external Wi-Fi interference. However, practical wireless networks rarely exhibit such favorable characteristics. It is, hence, desirable to design D-MIMO Wi-Fi networks which address the dynamic resource management challenges mentioned in Section~\ref{sec:introduction} (known to be NP-Hard), particularly when network conditions are dynamic. With this high-level goal in mind, the following discussion describes a few example scenarios to motivate the use of DRL to improve the performance of D-MIMO Wi-Fi networks. The following discussion assumes an enterprise D-MIMO Wi-Fi network, managed by an administrator, as the network of interest. 

Before studying these scenarios in detail, it is important to understand what is meant by the `performance' of a D-MIMO Wi-Fi network. There are many different quantities of interest in a wireless network, the suitability of which depend greatly on the target applications. This paper considers the throughput performance of users as the metric of interest; it could be average throughput of users or the throughput obtained by some percentile of users in the network. The specifics of the DRL framework as well as the different agents used will be explained in detail in Sections~\ref{sec:drl_framework} and \ref{sec:results} respectively; this section will keep the discussion to a generic DRL agent. 

\subsection{Vanilla Channel Assignment}
\label{ssec:P1}
\begin{figure*}[t]
	\centering
	\begin{minipage}[t]{0.3\textwidth}
		\centering
		\includegraphics[width=0.9\linewidth]{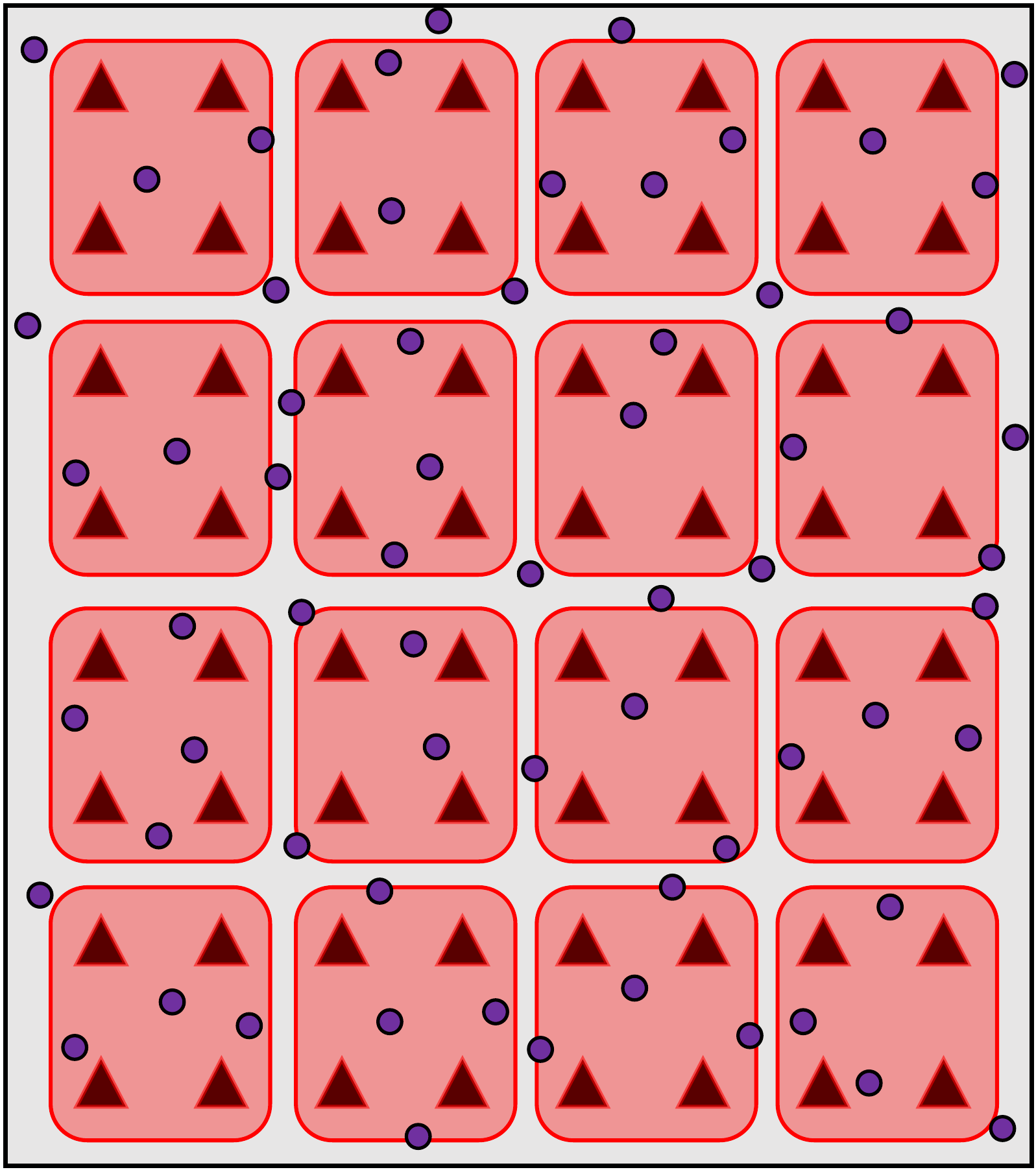}
		\captionof{figure}{A D-MIMO Wi-Fi network with $16$ groups (with four $\RH$s each), all assigned to the same channel. Triangles represent $\RH$s and circles represent users.}
		\label{fig:all_groups_same_channel}
	\end{minipage} \hfill
	\begin{minipage}[t]{0.3\textwidth}
		\centering
		\includegraphics[width=0.9\linewidth]{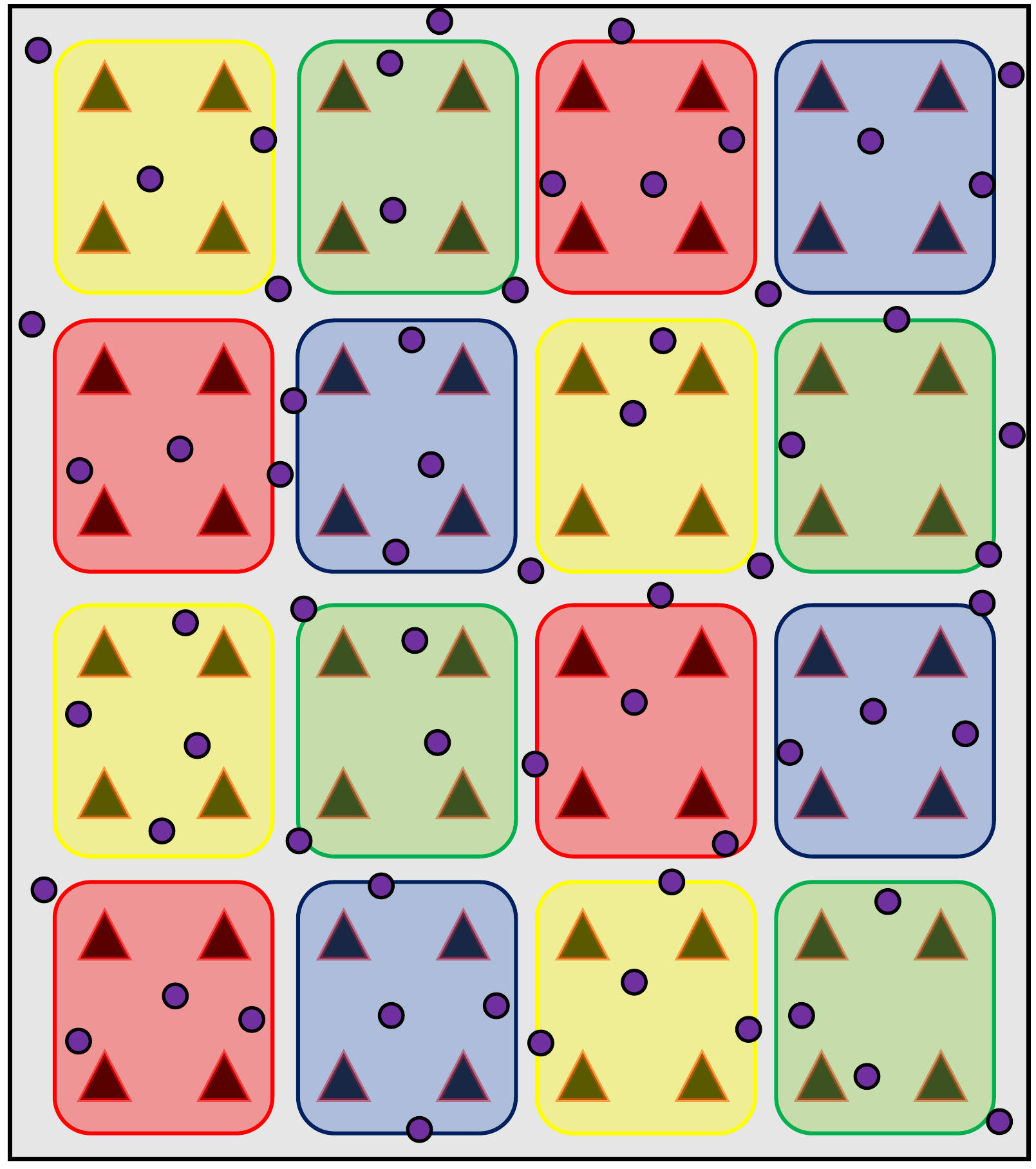}
		\captionof{figure}{Channel assignment based on a simple heuristic. Each color represents a unique non-overlapping channel. }
		\label{fig:groups_different_channels}
	\end{minipage} \hfill
	\begin{minipage}[t]{0.3\textwidth}
		\centering
		\includegraphics[width=0.9\linewidth]{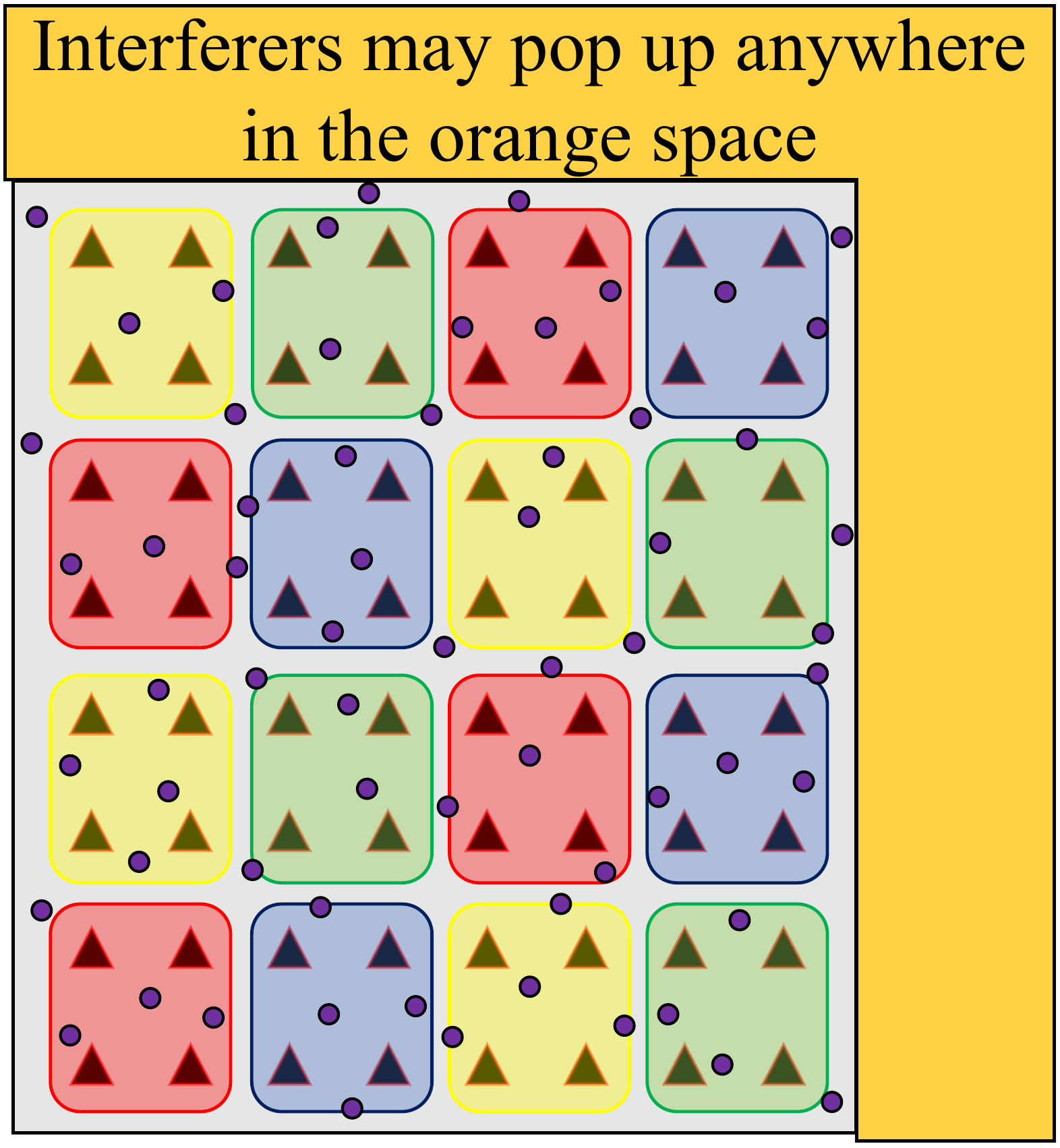}
		\captionof{figure}{A D-MIMO Wi-Fi network with random external Wi-Fi interference in its vicinity. The interferers may operate in channels red, blue, yellow or green.}
		\label{fig:dmimo_with_external_interference}
	\end{minipage}
\end{figure*}
First, consider a problem not specific just to D-MIMO networks but is common to most wireless networks -- channel assignment. Considering Wi-Fi networks in general, the goal is to determine which Wi-Fi AP gets assigned which of the available channels in order to maximize user throughput performance. In the particular case of D-MIMO, the goal will be slightly modified as determining the best group-channel assignment policy. Consider a D-MIMO Wi-Fi network with sixteen groups and four non-overlapping channels available. If the network starts with the worst channel assignment strategy (see Fig.~\ref{fig:all_groups_same_channel} in which all groups are assigned the same red channel), a DRL agent should determine how to assign the available channels to the D-MIMO groups in order to attain the best user throughput performance.

The channel assignment problem is chosen as a starting point because it is not restricted only to D-MIMO networks. It is easy to see that the D-MIMO network in Fig.~\ref{fig:all_groups_same_channel} is, in fact, just a typical Wi-Fi network deployment consisting of $16$ Wi-Fi APs, with each AP having spatially separated antennas (at the locations of the $\RH$s). Hence, the discussion is applicable to simple Wi-Fi networks not implementing D-MIMO as well.  

Determination of the optimal channel assignment policy is known to be NP-Hard \cite{mishra2005weighted}. For the simple network described in Fig.~\ref{fig:all_groups_same_channel}, channel assignment may be performed by simple heuristics like spacing out D-MIMO groups on the same channel far from each other in order to lower the channel contention among these groups as well as the interference from co-channel groups. The resulting channel assignment may look as shown in Fig.~\ref{fig:groups_different_channels}. If the performance of the DRL agent converges to what the heuristic assignment achieves, it suffices to say that the agent is effective in determining the best group-channel assignment policy. \emph{It is important to study the effectiveness of DRL agents in basic problems like the vanilla channel assignment before using them to approach more complex network scenarios.} 

\subsection{Channel Assignment with External Wi-Fi Interference}
\label{ssec:P2}
The next scenario of interest is to use a DRL agent to make a D-MIMO Wi-Fi network resilient to random external Wi-Fi interference (see Fig.~\ref{fig:dmimo_with_external_interference}). External Wi-Fi interferers may randomly appear in the orange zone in the vicinity of the considered D-MIMO network and each interferer may be assigned one of the four channels used by the D-MIMO groups. The goal of the DRL agent now is to update the channel assignment policy such that groups close to external interferers do not use the same channel as the latter. This will trigger changes to channel assignments of other groups which need not be neighbors of the external interferers. 

There exists extensive literature on channel assignment for APs in Wi-Fi networks based on heuristics, as described in \cite{chieochan2010channel}. Of these, HSUM is a popular heuristic algorithm based on weighted graph coloring, as described in \cite{mishra2005weighted}. Authors of the HSUM algorithm model channel assignment as a minimum-sum weighted vertex coloring problem in which different weights are put on interference edges. Looking at interference from the perspective of the users, this approach attempts to minimize the maximum interference as seen by clients in all common interfering regions. Specifically, in HSUM, Wi-Fi APs are required to transmit their interference metrics to their AP peers in order to facilitate a global view of the network topology at each AP. The maximum weighted interference is then minimized by HSUM in a global sense. Since the HSUM algorithm focuses mainly on Wi-Fi networks operating under the same administrative domain, we use HSUM to benchmark the performance of the DRL agent. 
\begin{figure*}[t] 
	\begin{subfigure}[t]{.48\linewidth}
		\centering
		\includegraphics[width=0.72\linewidth]{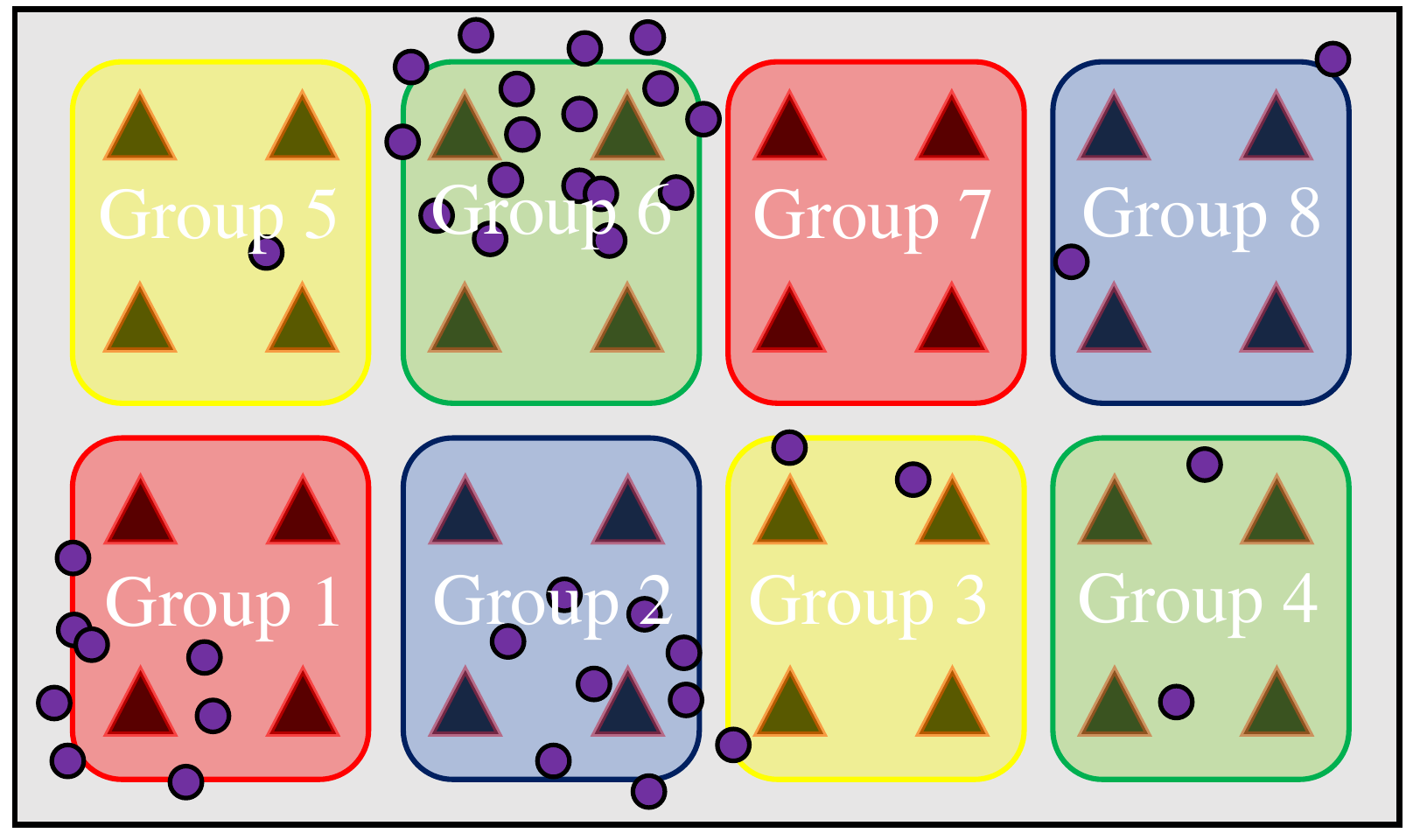}
		\caption{Grouping of neighboring $\RH$s; this arrangement is called adjacent grouping}
		\label{fig:non_uniform_distribution_adjacent_grouping}
	\end{subfigure} \hfill
	\begin{subfigure}[t]{.48\linewidth}
		\centering
		\includegraphics[width=0.72\linewidth]{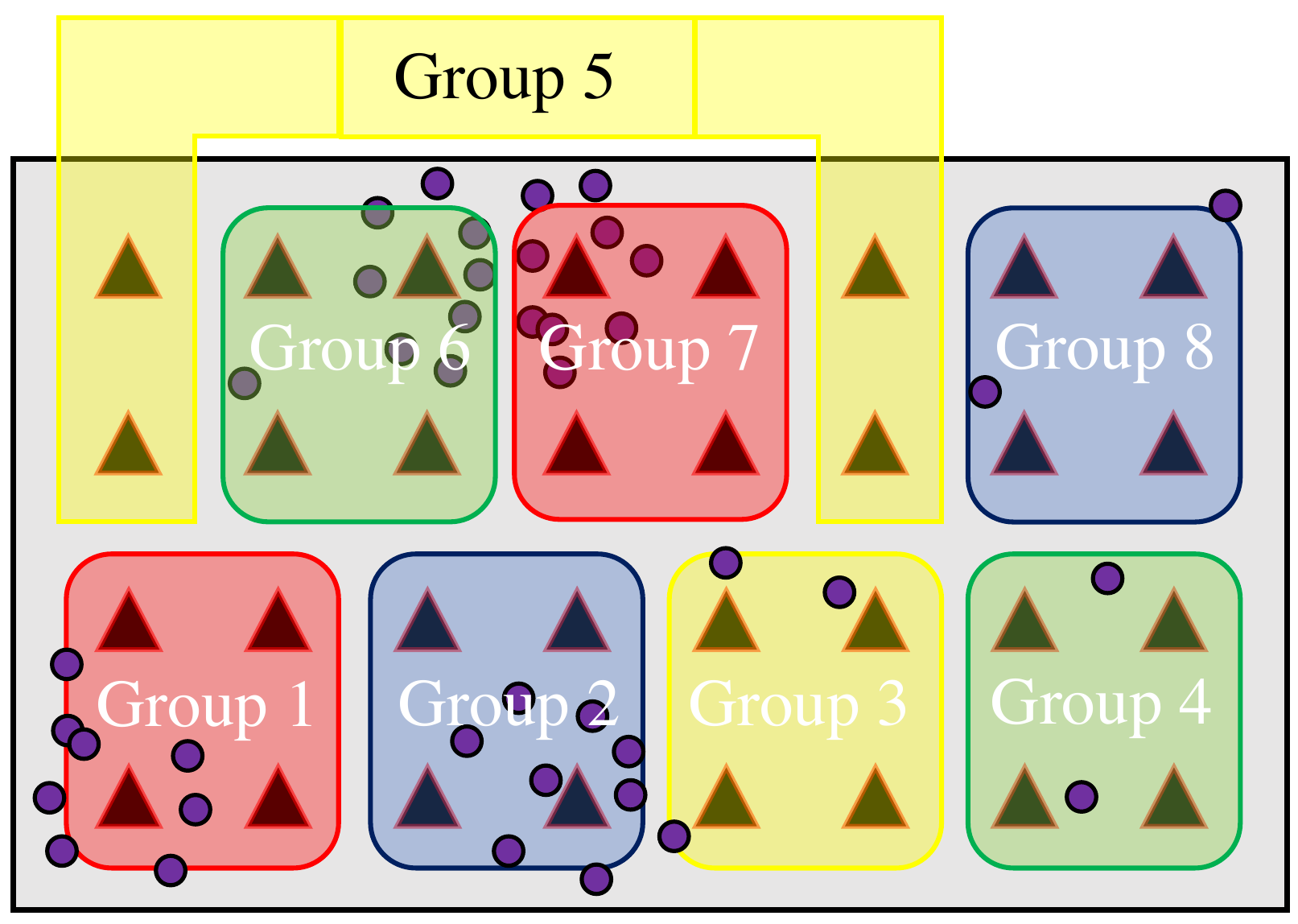}
		\caption{Updated grouping of $\RH$s in response to the user distribution}
		\label{fig:non_uniform_distribution_drl_grouping}
	\end{subfigure} \hfill
	
	\caption{A D-MIMO Wi-Fi network with $32$ $\RH$s (represented by triangles) and users (represented by circles) non-uniformly distributed in space. The arrangement in Fig.~\ref{fig:non_uniform_distribution_drl_grouping} achieves an improvement of $20\%$ in average user throughput compared to Fig.~\ref{fig:non_uniform_distribution_adjacent_grouping}.}
	\label{fig:DMIMO_non_uniform_distribution_of_users}
\end{figure*}
\subsection{Meeting Multiple Objectives}
\label{ssec:P3}
The scenarios considered so far consisted of a single objective to be met which was to maximize the throughput performance of users. This section, however, explores the use of a DRL agent to help a D-MIMO network meet multiple network objectives. Specifically, consider the case of channel assignment in the presence of random external Wi-Fi interference (see Fig.~\ref{fig:dmimo_with_external_interference}) but now the agent has two goals to meet: maximize the average throughput of users \textbf{\emph{and}} the fairness of throughput distribution among users. There is a separate branch in reinforcement learning literature studying this class of problems called multiple objective reinforcement learning (MORL). At a high level, MORL algorithms can be classified into different categories based on the following criteria: 
\begin{itemize}
    \item linear vs. non-linear scalarization of rewards from different objectives
    \item meeting different objectives with a single policy vs. multiple policies 
    \item training of the agent based on value iteration vs. policy iteration 
\end{itemize}
Section~\ref{ssec:P3_results} and the latter part of Section~\ref{sec:drl_framework} discuss the ideas of scalarization and policy vs. value iteration based training respectively in more detail. This is so because these notions will be more accessible after familiarization with the basic framework of DRL, which the initial part of Section~\ref{sec:drl_framework} provides. The objective of the DRL agent in this case is to update the channel assignments of D-MIMO groups in response to the external Wi-Fi interference such that both the average throughput of users as well as the fairness of throughput among the users is maximized. The yardstick for comparison of the performance of the DRL agent is again the HSUM-based channel assignment.

\subsection{D-MIMO RH grouping}
\label{ssec:P4}
The following discussion examines a problem specific to D-MIMO networks. The D-MIMO Wi-Fi network considered in Fig.~\ref{fig:groups_different_channels} is such that four neighboring $\RH$s form one group; this arrangement is referred to as \emph{adjacent grouping}. It is not clear if this arrangement is indeed the best $\RH$ clustering policy, particularly when users are non-uniformly distributed in the network space. Consider the scenario described in Fig.~\ref{fig:DMIMO_non_uniform_distribution_of_users} in which a D-MIMO network consists of $32$ $\RH$s. The colors in Fig.~\ref{fig:DMIMO_non_uniform_distribution_of_users} identify the channel assigned to the groups. Several users have been distributed non-uniformly in the networks space such that there is a dense concentration of users around a few $\RH$s. A practical example of such a user distribution is the presence of a conference or a meeting room in an office space where users typically congregate. The D-MIMO network should cater to higher data demands in this room compared to the rest of the office space. With adjacent grouping of $\RH$s (as shown in Fig.~\ref{fig:non_uniform_distribution_adjacent_grouping}), the dense concentration of users will have to be served by $\RH$s belonging to the same group (group 6 in Fig.~\ref{fig:non_uniform_distribution_adjacent_grouping}). However, if the $\RH$ clustering is modified to be the arrangement as shown in Fig.~\ref{fig:non_uniform_distribution_drl_grouping}, then the dense concentration of users gets split to be served by two groups (groups 6 and 7 in Fig.~\ref{fig:non_uniform_distribution_drl_grouping}) which, in fact, improves the average user throughput performance by up to $20\%$ (this result was obtained from simulations). It also helps that the rest of the office space is sparsely populated with users. Notice that $\RH$s belonging to group 5 and group 3 are assigned channel yellow and they are close to each other. However, this does not become a complication since $\RH$s in group 5 do not have any users to serve and hence do not contend with group 3 for access of channel yellow. Note that the arrangement in Fig.~\ref{fig:non_uniform_distribution_drl_grouping} may not be the optimal grouping for that user distribution; this example is used to illustrate the importance of clustering of $\RH$s in improving user throughput performance. The goal of the DRL agent, in this scenario, is to update the $\RH$ grouping arrangement in response to the distribution of users in the network space. 

%% file: drl_framework.tex
\section{Deep Reinforcement Learning Framework}
\label{sec:drl_framework}
\begin{figure}[t]
\centering
    \includegraphics[width=0.95\linewidth]{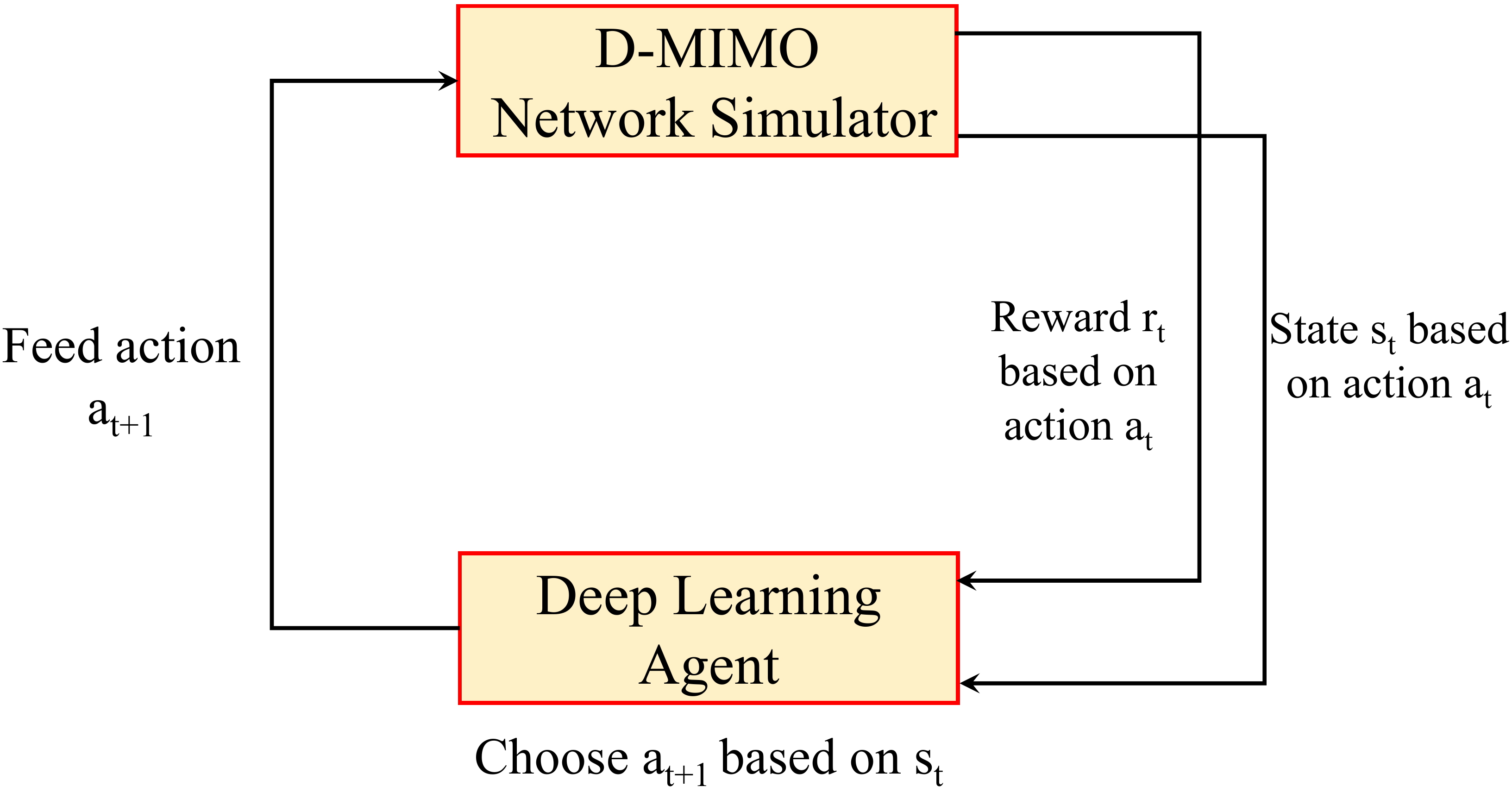}
    \caption{Deep reinforcement learning framework}
    \label{fig:drl_framework}
\end{figure}

Fig.~\ref{fig:drl_framework} describes the high-level architecture of the DRL framework implemented in this paper. The framework consists of the custom D-MIMO network simulator introduced in \cite{neel2018dmimoo} along with a learning agent. The agent, modeled as a \emph{deep neural network}, learns about the environment it is acting on by carrying out different actions ($a$) based on the current state ($s$) of the environment, and the reward ($r$) it receives for each carried out action; the agent receives a positive reward if the action it chose resulted in a better performance of the network compared to before, else it receives a negative reward. Before delving any further into the specifics of DRL, this section introduces some associated terminology. Let $\mathcal{A}$ denote the set of all possible actions that could be performed by the agent, and $\mathcal{S}$ denote the set of all states that the environment could be in. The subscript $t$ in the following discussion denotes the time/step index at which the agent interacts with the environment. 
\begin{itemize}
    \item \emph{Environment}: The D-MIMO Wi-Fi network of interest serves as the environment on which the deep learning agent acts. The environment can be modeled as a markov decision process (MDP) with a state transition matrix, of dimension $|\mathcal{A}| \times |\mathcal{S}|$, which provides the probability $p(s_{t+1} | s_t,a_t)$ of moving from state $s_t$ to $s_{t+1}$ when action $a_t$ is performed. An MDP is also characterized by the reward model which describes the real-valued reward value that the agent receives for choosing action $a$ in state $s$. The environment feeds the state information and the obtained reward for an action to the agent. 
    \item \emph{Episode}: An episode is a sequence of actions carried out by the agent, and the corresponding states and rewards obtained from the environment. An episode terminates with either a terminal state or when a certain number of actions have been carried out. Typically, games are episodic wherein an episode completes when a player has won or lost the game. In the scenarios considered in this paper, however, \emph{one learning episode is terminated when the number of carried out actions exceeds a threshold $T$}, that is, an episode consists of a fixed number of actions. This is a deliberate choice because the object of the agent is to maximize/improve the throughput performance of the D-MIMO network and hence there exists no clear winning or losing condition. One may advocate for a threshold based episode termination in which an episode is declared to be won if the throughput metric exceeds a predetermined threshold. That may work, but the DRL agents implemented in this paper are free to maximize the performance without any hard constraints imposed on them. 
    \item \emph{Policy}: The actions of an agent are governed by a map called the policy, denoted by $\pi$. It decides the probability of the agent choosing an action $a_t=a$ when the environment is in state $s_t=s$. A policy is usually parameterized by parameters $\theta$. The mathematical definition of a policy is provided by (\ref{eqn:policy}).
    \begin{equation} 
    \label{eqn:policy}
    \begin{split}
        \pi &: \mathcal{S} \times \mathcal{A} \rightarrow [0,1] \\
        \pi_\theta(a|s) &= \mathbb{P}\left[a_t = a\,\given[\Big]\, s_t = s\right]
        \end{split}
    \end{equation}
    Note that policy $\pi_\theta$ is stochastic, that is, $\pi_\theta(a|s)$ is modeled as a probability distribution over the set of all actions $\mathcal{A}$ given the current state of the environment.
    \item \emph{State-value function}: Value function quantifies how good it is to be in a given state `$s$'. It is defined as the expected return starting at state $s_t$ following a policy $\pi$. Note that the reward obtained for choosing action $a_t=a$ when at state $s_t=s$ is represented by $r_t$. 
    \begin{equation}
        \label{eqn:state_value_function}
        V^\pi(s) = \mathbb{E}[R] = \mathbb{E}\left[\sum\limits_{k=0}^{T-t}\gamma^kr_{t+k}\,\given[\Big]\,s_t = s\right],
    \end{equation}
    where R denotes the cumulative discounted return and $\gamma$ denotes the discount factor. 
    \item \emph{Discount factor}: Denoted by $\gamma$, discount factor was originally conceived as a mathematical trick, in case of non-episodic environments where the number of actions $|\mathcal{A}|\rightarrow\infty$, to make the infinite sum in (\ref{eqn:state_value_function}) (when $T\rightarrow\infty$) finite. The value of the discount factor is bounded between $[0,1]$ and it determines the importance of the future rewards. $\gamma$ closer to $0$ makes the agent opportunistic by considering rewards in the immediate future whereas $\gamma$ close to $1$ prioritizes rewards in the distant future. 
\end{itemize}

\begin{figure*}[t] 
	\captionsetup[subfigure]{justification=centering}
	\begin{subfigure}[t]{.24\linewidth}
		\centering
		\includegraphics[width=\linewidth]{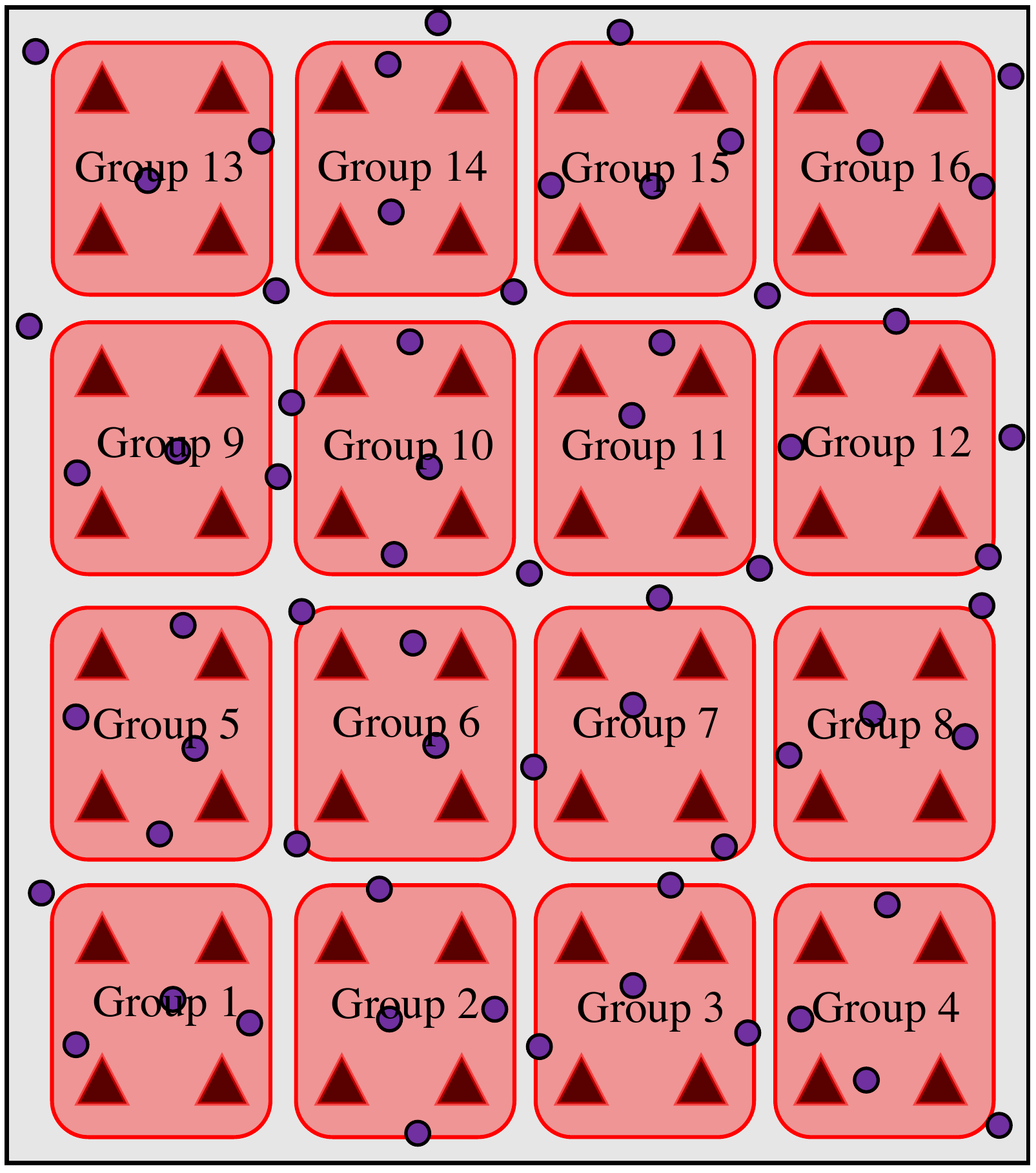}
		\caption{Step $t_0$ \\ Initial state $s_0$}
	\end{subfigure} \hfill
	\begin{subfigure}[t]{.24\linewidth}
		\centering
		\includegraphics[width=\linewidth]{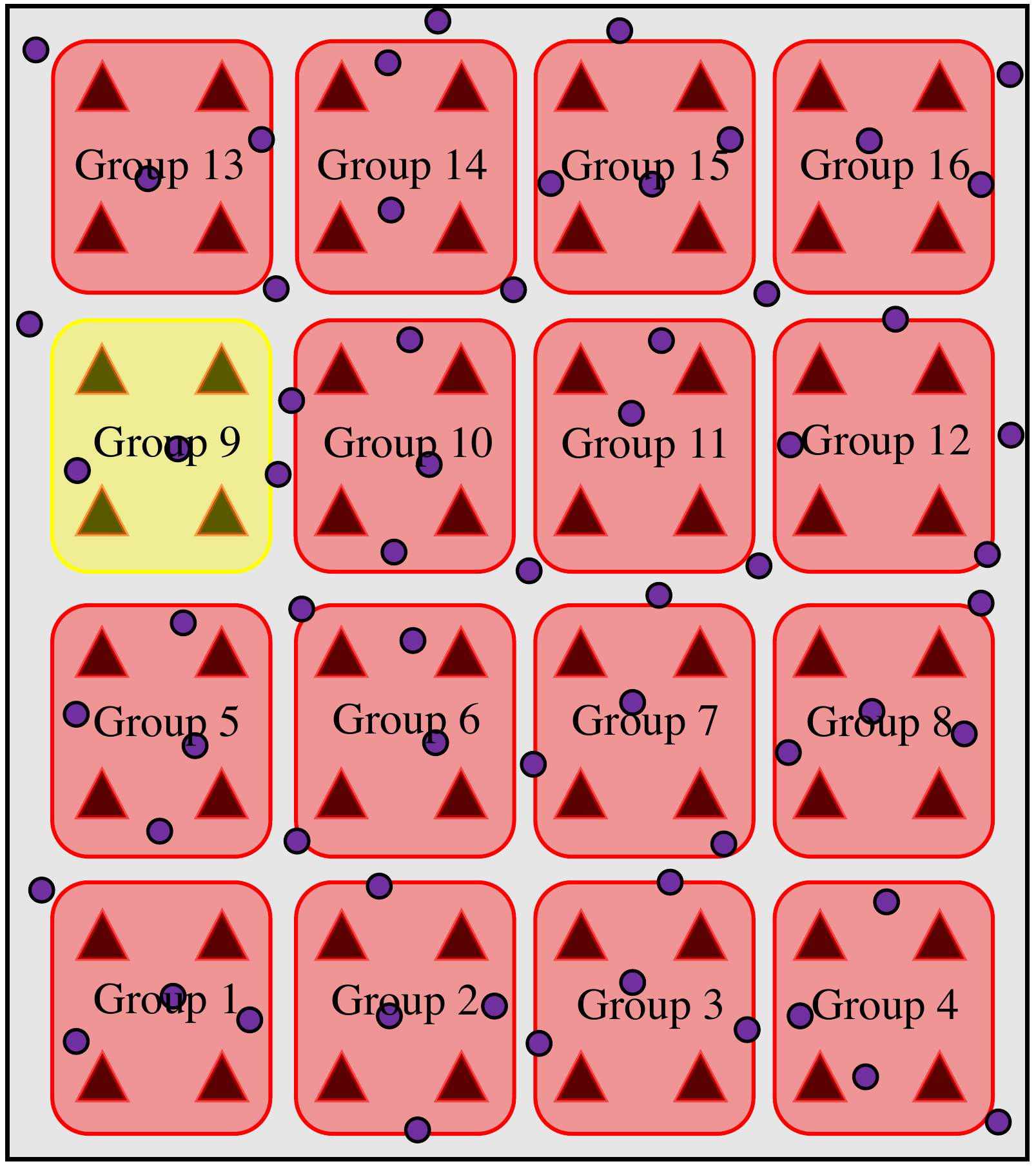}
		\caption{Step $t_1$ \\ Assign channel yellow to group 9 based on state $s_0$}
	\end{subfigure} \hfill
	\begin{subfigure}[t]{.24\linewidth}
		\centering
		\includegraphics[width=\linewidth]{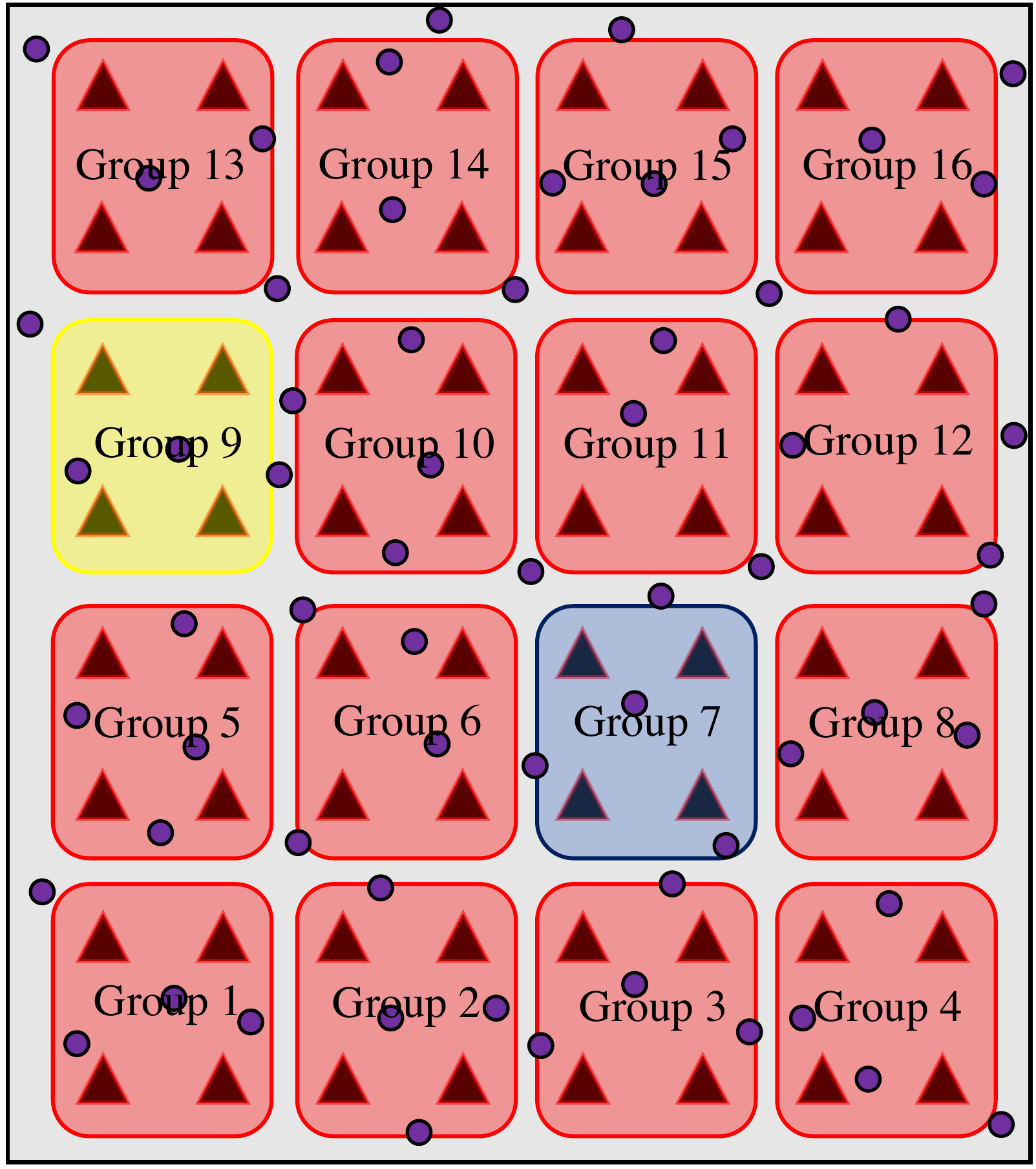}
		\caption{Step $t_2$ \\ Assign channel blue to group 7 based on state $s_1$}
	\end{subfigure} \hfill
	\begin{subfigure}[t]{.24\linewidth}
		\centering
		\includegraphics[width=\linewidth]{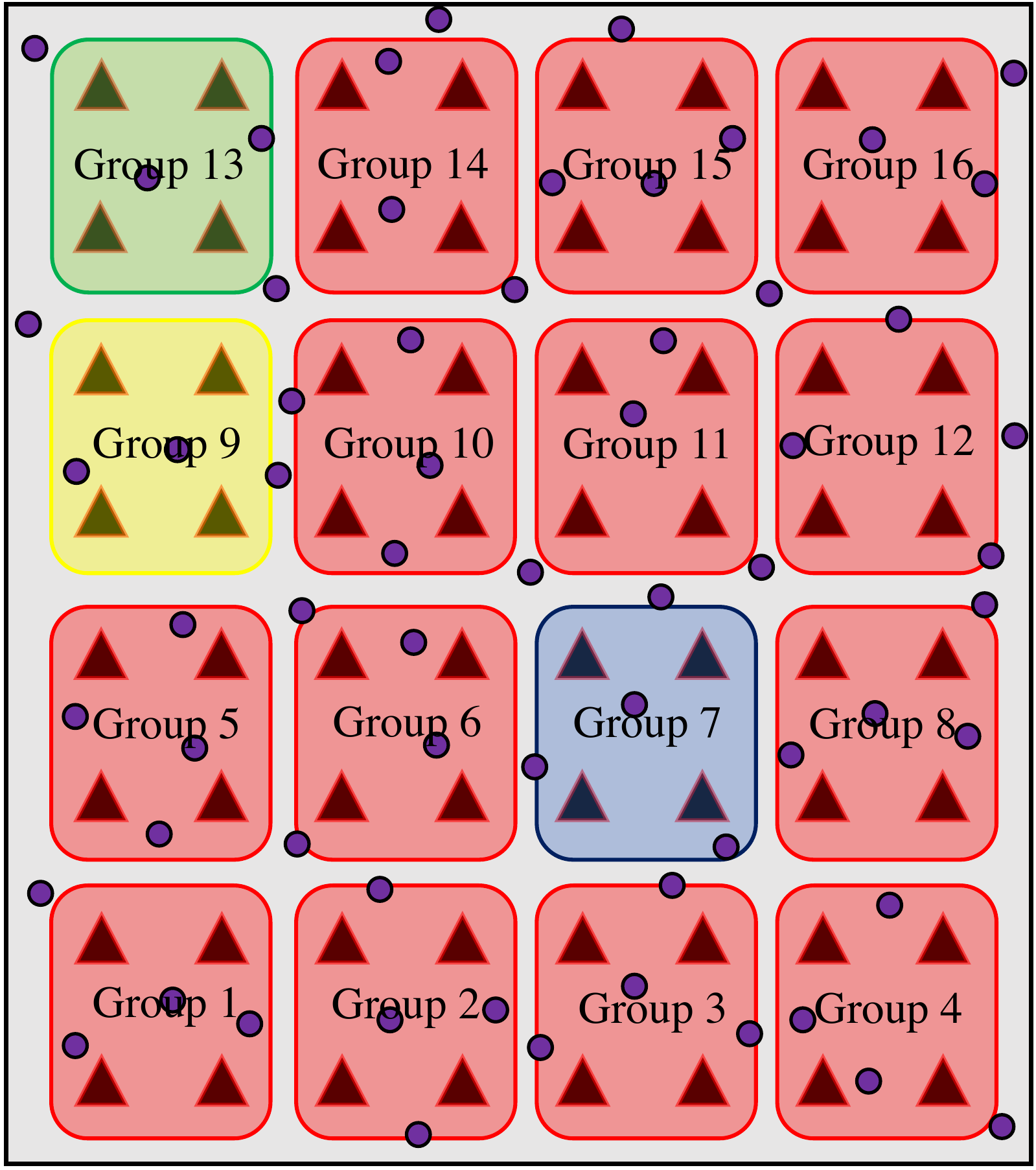}
		\caption{Step $t_3$ \\ Assign channel green to group 13 based on state $s_2$}
	\end{subfigure} \hfill
	
	\caption{Timeline of how a typical episode (in case of the channel assignment problem) progresses. $\RH$s are represented by triangles whereas users are represented by circles.}
	\label{fig:timeline_of_an_episode}
\end{figure*}

It is worthwhile to digress a bit to clarify the idea of actions, states and rewards in the context of a D-MIMO Wi-Fi network. Consider the timeline of an example learning episode as described in Fig.~\ref{fig:timeline_of_an_episode} in case of the simple channel assignment problem for D-MIMO discussed in Section~\ref{ssec:P1}. The considered D-MIMO network consists of sixteen groups with four $\RH$s each, and four non-overlapping available channels. An episode begins with the D-MIMO network in its initial state $s_0$. \textbf{State}, at any step, is the current group-channel assignments. For the network shown in Fig.~\ref{fig:timeline_of_an_episode}, state at step $t$ (denoted by $s_t$) is a $16\times1$ vector $[c_1,c_2,c_3,....,c_{16}]^T$ where $c_i$ represents the channel assigned to group $i$ and $[.]^T$ denotes the transpose operator. Let the throughput metric in the initial state be $x_0$. At step $t_1$, the agent chooses \textbf{action} $a_1$, based on the initial state $s_0$, which is to assign channel yellow to group 9. The D-MIMO network simulator receives this action, performs the necessary channel assignment update, and simulates the network to obtain the throughput metric $x_1$. The simulator computes the difference between the throughput metrics $x_1$ and $x_0$ as the \textbf{reward} $r_1$ for action $a_1$. The simulator feeds state $s_1$ and reward $r_1$ to the agent. Based on the updated state $s_1$, the agent chooses action $a_2$ at step $t_2$, which in this case is to assign channel blue to group 7. The environment then computes the new reward $r_2$ and passes it, along with the updated state $s_2$, to the agent. This cycle continues until the number of actions chosen exceeds the threshold $T$; this marks the end of one episode. Once an episode terminates, the environment resets back to its initial state $s_0$, that is, all groups are assigned channel red, and the timeline continues. After the completion of an episode, the agent computes the cumulative discounted return (as shown in (\ref{eqn:state_value_function})) which is then used to update the weights of its neural network depending on the training algorithm of choice. 

Note that state, action, and reward information change when considering problems other than the simple channel assignment problem. The aforementioned discussion is intended to be an example to better understand these concepts and see how an episode progresses with time. Specific details regarding the choice of state, action, and reward will be discussed individually for each problem in Section~\ref{sec:results}. 

Coming back to reinforcement learning terminology, the objective of an agent is to determine the best policy $\pi$ which will maximize the value function given in (\ref{eqn:state_value_function}). The optimal policy $\pi^*$ can be formulated as follows:
\begin{equation}
    \pi^* = \argmax_\pi \, V^\pi(s), \, \forall s \in \mathcal{S} 
\end{equation}
The optimal policy $\pi^*$ is optimal for all states $s$, that is, the same policy can be used regardless of what the initial state of the environment is. Broadly, reinforcement learning algorithms are classified into two categories based on how to they arrive at the optimal policy $\pi^*$: \emph{policy iteration} in which policies are directly searched, evaluated and improved, and \emph{value iteration} in which the optimal value function is first determined from which the optimal policy is extracted. Policy iteration methods converge quicker compared to value iteration and have been more commonly used when the action space of the environment is large, for example, continuous action space in robotics. Because there is an infinite number of actions and/or states to estimate the values for, value-based approaches are very highly computationally expensive since these methods involve computing the value function in (\ref{eqn:state_value_function}) for all states. Policy gradient methods were used by DeepMind for playing the game AlphaGo which has a large discrete action space \cite{silver2016mastering,silver2017mastering}. This paper will also use policy gradients to train the agent since the action space in case of D-MIMO Wi-Fi networks is large. For instance, in case of the simple channel assignment problem described in Section~\ref{ssec:P1}, the number of actions that the agent can potentially take in each step is $64$, since there are $16$ groups and $4$ available channels. If one episode consists of $T$ steps, then the number of possible action combinations is $64^T$ which is exponentially large. 

\subsection{Policy Gradients}
\label{sec:policy-gradients}
As discussed before, policy $\pi$ is modeled as a parameterized function with respect to $\theta$. Policy gradient methods aim at modeling and optimizing the policy directly. Let the reward function be redefined as follows:
\begin{equation}
\label{eqn:reward_function}
    J(\theta) = \sum\limits_{s\in\mathcal{S}} d^\pi(s)V^\pi(s) = \sum\limits_{s\in\mathcal{S}}d^\pi(s)\sum\limits_{a\in\mathcal{A}}\pi_\theta(a|s)Q^\pi(s,a),
\end{equation}
where $d^\pi(s)$ represents the stationary distribution of Markov chain for $\pi_\theta$, and $Q^\pi(s,a)$ denotes the state-action pair value function which is similar to the value function described in (\ref{eqn:state_value_function}) and is defined as $Q^\pi(s,a) = \mathbb{E}\left[\sum\limits_{k=0}^{T-t}\gamma^kr_{t+k}|s_t=s,a_t=a\right]$, i.e, it describes the value of a state-action pair when the agent follows the policy $\pi$. Since the objective of the DRL agent is to maximize its reward function, the parameters $\theta$ of the policy need to be moved in the direction of the gradient of the reward function to find the best $\theta$ that produces the highest return, that is, 
\begin{equation}
    \theta \leftarrow \theta + \alpha\nabla_\theta J(\theta)
\end{equation}
where $\alpha$ denotes the learning rate of the agent. 

At first glance, computing the gradient of the reward function in (\ref{eqn:reward_function}) might seem difficult because the reward function depends both the action selection as well as the stationary distribution of states, both of which depend on $\theta$ directly or indirectly. However, the policy gradient theorem (described in \cite{sutton2018reinforcement, policygradients}) provides a formulation for the gradient which does not involve the derivative of the state distribution $d^\pi(s)$ and the gradient can be computed as 
\begin{equation}
\label{eqn:grad_reward_function}
        \nabla_\theta J(\theta) = \mathbb{E}_\pi[Q^\pi(s,a)\nabla_\theta \ln(\pi_\theta(a|s))],
\end{equation}
where $\mathbb{E}_\pi$ is a simplified notation for $\mathbb{E}_{s\sim d^\pi, a\sim \pi_\theta}$, that is, both state and action distributions follow the policy $\pi_\theta$. Such algorithms are called \emph{on-policy} algorithms in which training samples are collected according to the policy the agent is optimizing for. Algorithms which use a different policy to sample training data are called \emph{off-policy} algorithms. 

There exists extensive literature on different kinds of policy gradient algorithms and the following discussion will focus on two such algorithms used in this work: REINFORCE agent \cite{sutton2000policy} and deep deterministic policy gradients \cite{lillicrap2015continuous}. 

\subsubsection{REINFORCE Agent}
\label{sssec:REINFORCE}
REINFORCE or the Monte-Carlo policy gradient \cite{sutton2000policy} relies on an estimated return by Monte-Carlo methods using episode samples to update the policy parameter $\theta$. The REINFORCE agent makes use of the fact that expectation of the sample gradient is equal to the actual gradient. The parameter update is described in the following procedure:
\begin{enumerate}
    \item Randomly initialize the parameters $\theta$ of the policy $\pi_\theta$ that the agent is optimizing for.
    \item Obtain an episode of length $T$ which consists of a sequence of state-action-reward-state-action (SARSA). The full sequence of an episode is called a trajectory. \\
    Trajectory $\tau = \{s_0,a_1,r_1,s_1,a_2, ... ,s_{T-1},a_T,r_T,s_T\}$
    \item For $t=1,2,...,T$
    \begin{itemize}
        \item Compute the cumulative discounted return $G_t = \sum\limits_{k=0}^{T-t}\gamma^kr_{t+k}$
        \item Update the parameters of the policy as $\theta \leftarrow \theta +\alpha\gamma^tG_t\nabla_\theta\ln\pi_\theta(a_t|s_t)$
    \end{itemize}
\end{enumerate}
This agent is \emph{on-policy} as it computes the cumulative discounted returns from sample episodes collected according to the policy $\pi_\theta$ and use them to update parameters $\theta$ of the same policy. This algorithm requires the full trajectory of an episode and hence is called a Monte-Carlo method. 

\subsubsection{Deep Deterministic Policy Gradients (DDPG)}
This is an \emph{off-policy deterministic actor-critic} algorithm. The following discussion discusses each of these terms individually. Off-policy algorithms were introduced in the introductory part of this section. There are major advantages to using \emph{off-policy} algorithms compared to on-policy:
\begin{itemize}
    \item Off-policy algorithms do not require full episodes and can reuse any past experiences for better sample efficiency. This is called `experience replay' through which the agent randomly samples from past stored state-action-reward-next state experiences; these experiences need not be part of the same episode. 
    \item The agent uses a different policy for sample collection than the target policy, which leads to better exploration. 
\end{itemize}
Throughout the discussion in this section till now, the policy $\pi_\theta(a|s)$ has been modeled as a probability distribution over the set of all actions $\mathcal{A}$ given the current state of the environment and hence it is stochastic. In \emph{deterministic policy gradients}, the policy is modeled as deterministic, that is, an action is a deterministic function of the current state ($a = \mu_\theta(s)$). The policy $\mu$ is again parameterized by parameters $\theta$. 

Use of deterministic policies necessitates a reformulation of the reward function definition. Since this algorithm is off-policy, let the training trajectories (sequence of state-action-reward-next state) be collected according to a stochastic policy $\beta(a|s)$. Since $\mu_\theta(s)$ is the policy the agent is attempting to learn, let it be known as the target policy. Let $\rho_0(s)$ denote the initial state distribution, $\rho^\beta(s\rightarrow s',k)$ denote the visitation probability density at state $s'$ after moving $k$ steps from state $s$ following the policy $\beta$, and $\rho^\beta(s') = \int_\mathcal{S} \sum\limits_{k=1}^{\infty}\gamma^{k-1}\rho_0(s)\rho^\beta(s\rightarrow s',k)ds$ represent the discounted state distribution. Then, the reward function (similar to the reward function in  (\ref{eqn:reward_function})) is redefined as 
\begin{equation}
\label{eqn:reward_function-deterministic}
    J(\theta) = \int_\mathcal{S}\rho^\beta(s)Q(s,\mu_\theta(s))ds. 
\end{equation}
Computing the gradient of the reward function in (\ref{eqn:reward_function-deterministic}) with respect to parameters $\theta$ using chain rule yields
\begin{equation}
\label{eqn:gradient_reward_function_deterministic}
\begin{split}
    \nabla_\theta J(\theta) &= \int_\mathcal{S}\rho^\beta(s)Q(s,a)\nabla_\theta \mu_\theta(s) |_{a = \mu_\theta(s)}ds \\
    &= \mathbb{E}_{s\sim\rho^\beta}\left[\nabla_a Q^\mu(s,a)\nabla_\theta \mu_\theta(s)|_{a = \mu_\theta(s)}\right].
\end{split}
\end{equation}
The computation of the gradient in (\ref{eqn:gradient_reward_function_deterministic}) requires taking an expectation over the state space only. Comparing this with the gradient in (\ref{eqn:grad_reward_function}), it is evident that the same gradient requires performing an expectation over both state and action spaces for the stochastic policy case, thus necessitating collecting more samples. Deterministic policy gradients are hence highly helpful when the action space is vast. In fact, this algorithm was originally conceived to be used in environments with continuous action spaces like robotics. Another major advantage of using deterministic policy gradients is the avoidance of importance sampling. 

DDPG algorithm belongs to the class of \emph{Actor--Critic methods} which attempt to learn the value function assisting the policy update in addition to the policy itself. The \emph{critic} updates the parameters of the value function (it could be either the state value function $V^\pi(s)$ or the state-action value function $Q^\pi(s,a)$) and the \emph{actor} updates the parameters $\theta$ of the policy $\pi_\theta$ in the direction suggested by the critic. 

This paper uses an extension of the DDPG algorithm to be applied in large discrete action spaces. Specifically, this work uses the Wolpertinger agent described in \cite{dulac2015deep} to reduce the size of the action space, particularly for the problem of $\RH$ grouping (described in Section~\ref{ssec:P4}) which has a vast action space. This agent avoids the heavy cost of evaluating all actions while retaining generalization over actions. The Wolpertinger architecture consists of three main parts: \emph{an actor network, K-Nearest Neighbors (K-NN), and a critic network}. The actor network reasons over actions within a continuous space and maps this output to a discrete action. The critic network is to correct the decision made by the actor network. The DDPG algorithm is applied to update both critic and actor networks. K-NN can help to explore a set of actions to avoid poor decisions. The reader is referred to \cite{dulac2015deep} for details regarding the implementation of the Wolpertinger agent.

%% file: results.tex
\section{Results}
\label{sec:results}
\begin{table}[t]
	\centering
	\caption{Specifics of the learning agent used in Sections~\ref{ssec:P1_results}, \ref{ssec:P2_results}, and \ref{ssec:P3_results}}
	\label{tab:specifics_of_REINFORCE_agent}
	\begin{tabular}{@{}ll@{}}
		\toprule
		Number of hidden layers & $1$ \\
		Number of input nodes & $16$ \\
		Number of hidden nodes & $48$ \\
		Number of output nodes & $64$ \\
		Configuration of layers & Densely connected \\
		Hidden node activation & Rectified linear unit (ReLU) \\ 
		Output node activation & Softmax \\
		Learning rate of the agent & $0.003$ \\
		Optimizer & Adam \cite{kingma2014adam}\\
		\bottomrule
	\end{tabular}
\end{table}
\begin{figure*}[t] 
	\begin{subfigure} {0.49\linewidth}
		\centering
		\includegraphics[width=\linewidth]{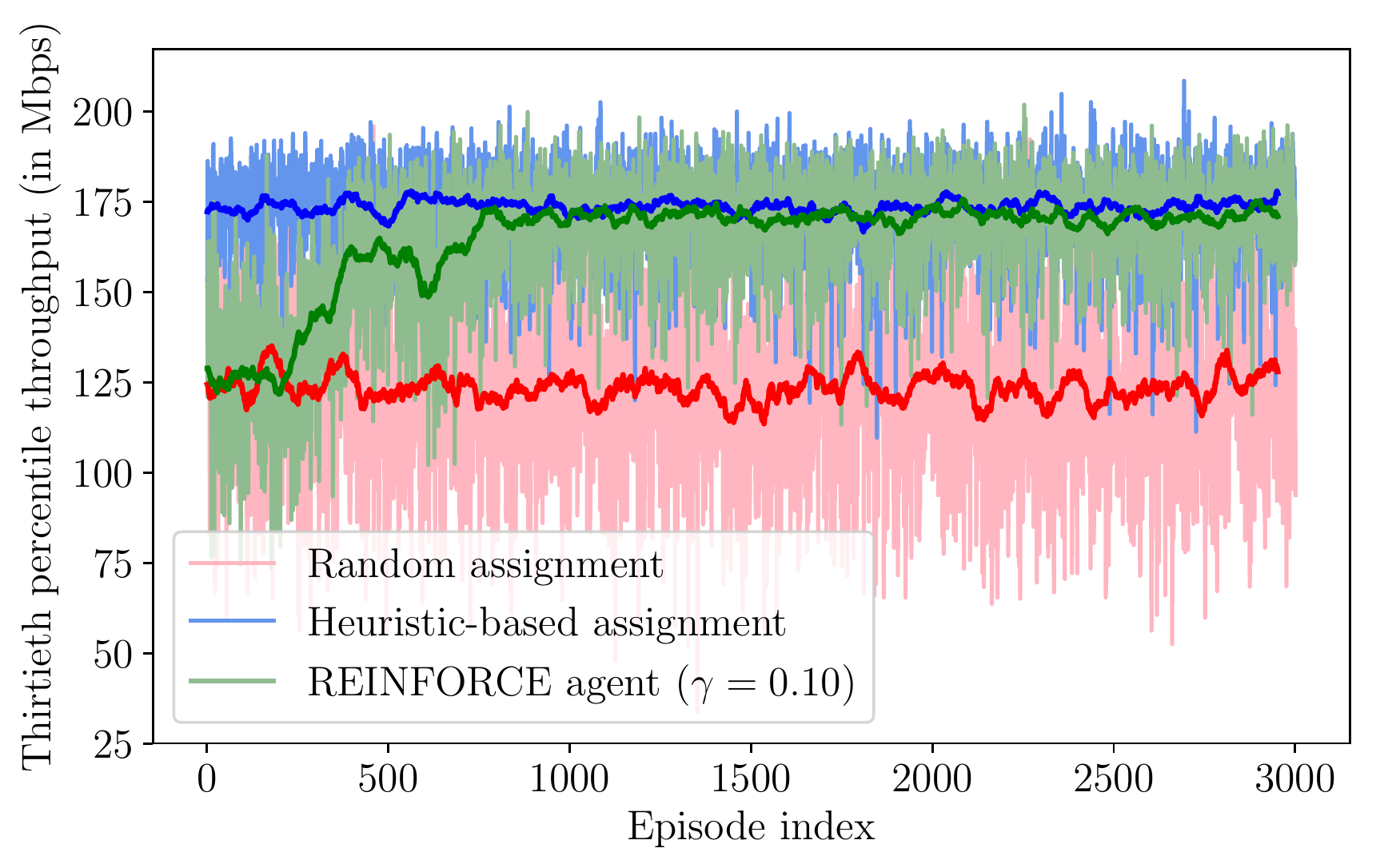}
		\caption{DRL Agent with $\gamma = 0.10$} 
		\label{fig:P1_gamma_10} 
	\end{subfigure} \hfill
	\begin{subfigure} {0.49\linewidth} 
		\centering
		\includegraphics[width=\linewidth]{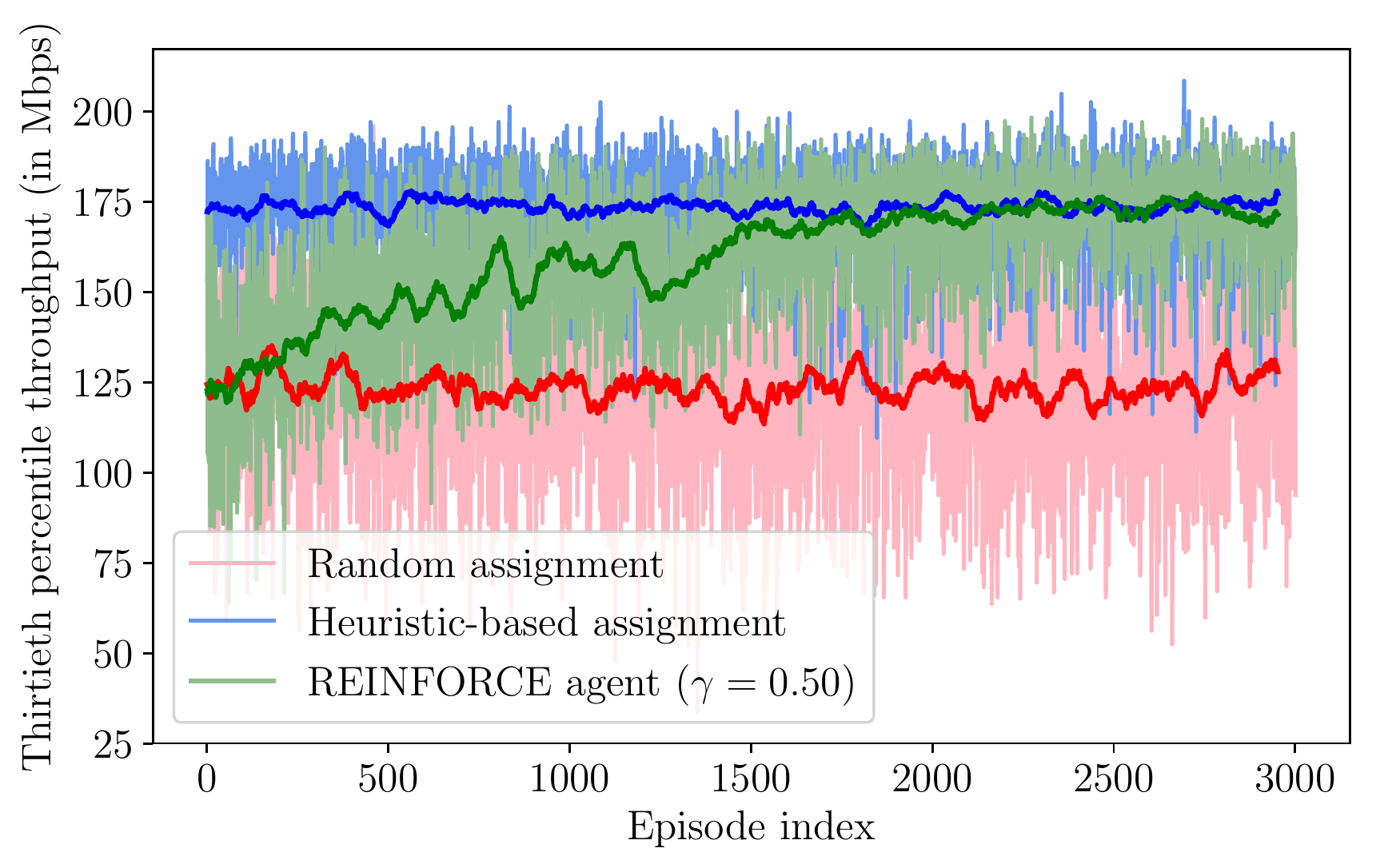}
		\caption{DRL Agent with $\gamma = 0.50$} 
		\label{fig:P1_gamma_50} 
	\end{subfigure} 
	\begin{subfigure}{1\textwidth}
		\centering
		\includegraphics[scale=0.49]{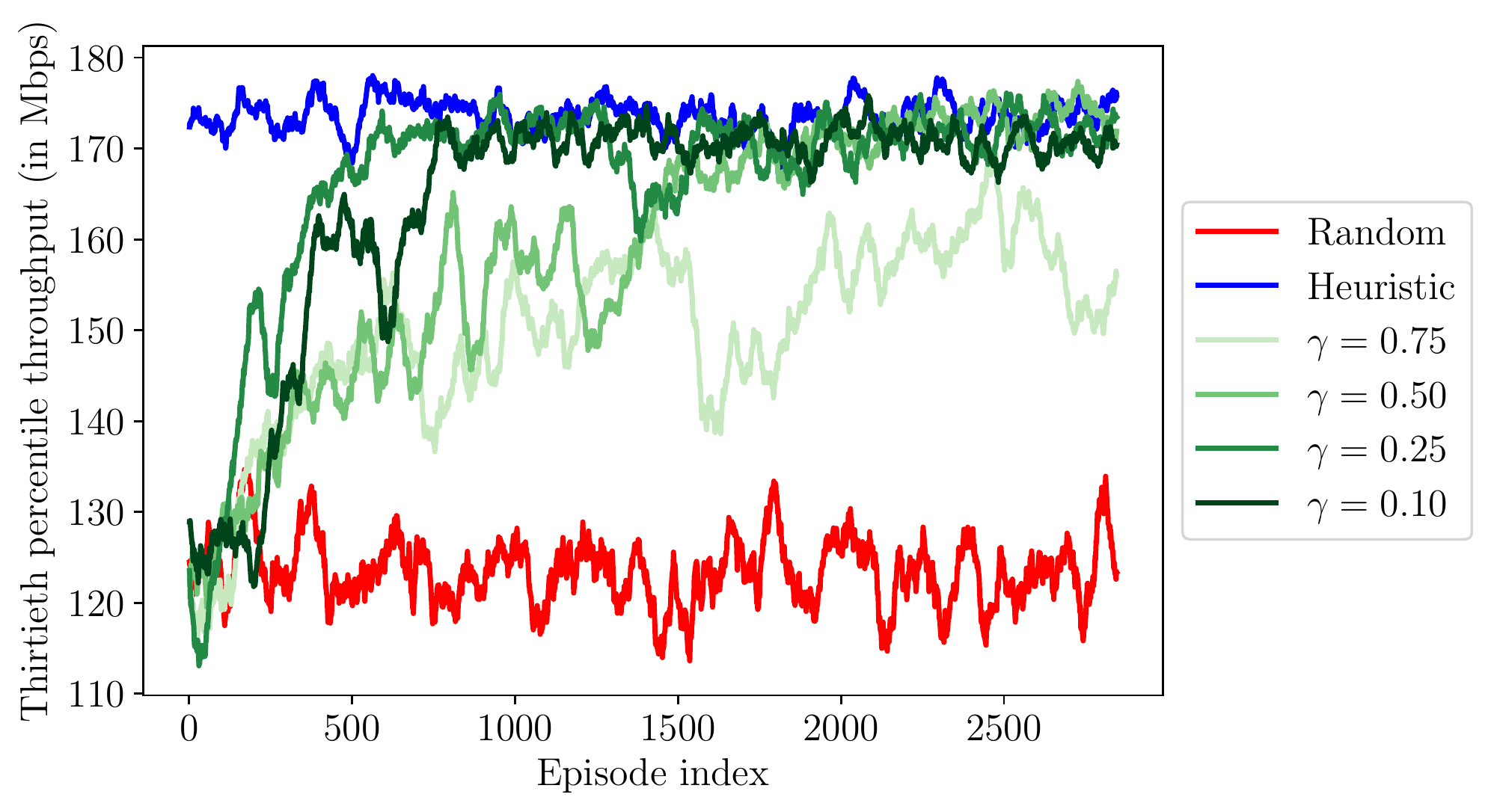} 
		\caption{Comparison of REINFORCE agents with different discount factors (moving average over $50$ episodes)} 
		\label{fig:P1_different_gamma} 
	\end{subfigure} 
	\caption{Throughput of the thirtieth percentile of users with different channel assignment schemes}
	\label{fig:P1_results}
\end{figure*}
This section revisits the problems described in Section~\ref{sec:motivating_scenarios}, discusses the specifics of the DRL agent used to approach the problems, and studies the results obtained from on-line DRL training. The DRL framework, shown in Fig.~\ref{fig:drl_framework}, was implemented using a combination of \textbf{OpenAI Gym} \cite{brockman2016openai} for the environment and \textbf{TensorFlow} \cite{tensorflow2015whitepaper} for the deep learning agent. OpenAI Gym was used as a wrapper outside the custom D-MIMO Wi-Fi simulator. For problems described in Sections~\ref{ssec:P1}, \ref{ssec:P2}, and \ref{ssec:P3}, the D-MIMO network of interest was an office floor of dimension \SI{80}{m} $\times$ \SI{80}{m} with $16$ D-MIMO groups (with four $\RH$s per group) and $64$ users uniformly distributed throughout the office space (see Fig.~\ref{fig:all_groups_same_channel}). The $\RH$s were separated (in $x$ and $y$ directions) by \SI{10}{m}. There were four non-overlapping channels, each of bandwidth \SI{80}{\MHz}, assumed to be available in the \SI{5}{\GHz} band. Other simulation parameters and the channel model used in the simulations can be found in \cite{neel2018dmimoo}. Each step in a DRL episode involved running a simulation of the D-MIMO network for a network time of \SI{100}{\ms}. The scenarios considered in this paper were exclusively downlink with full buffer traffic to all users. Note that learning was episodic and the number of actions per episode was arbitrarily chosen to be fifty.
\begin{figure*}[t] 
	\begin{subfigure}[t]{0.49\linewidth} 
		\centering
		\includegraphics[width=\linewidth]{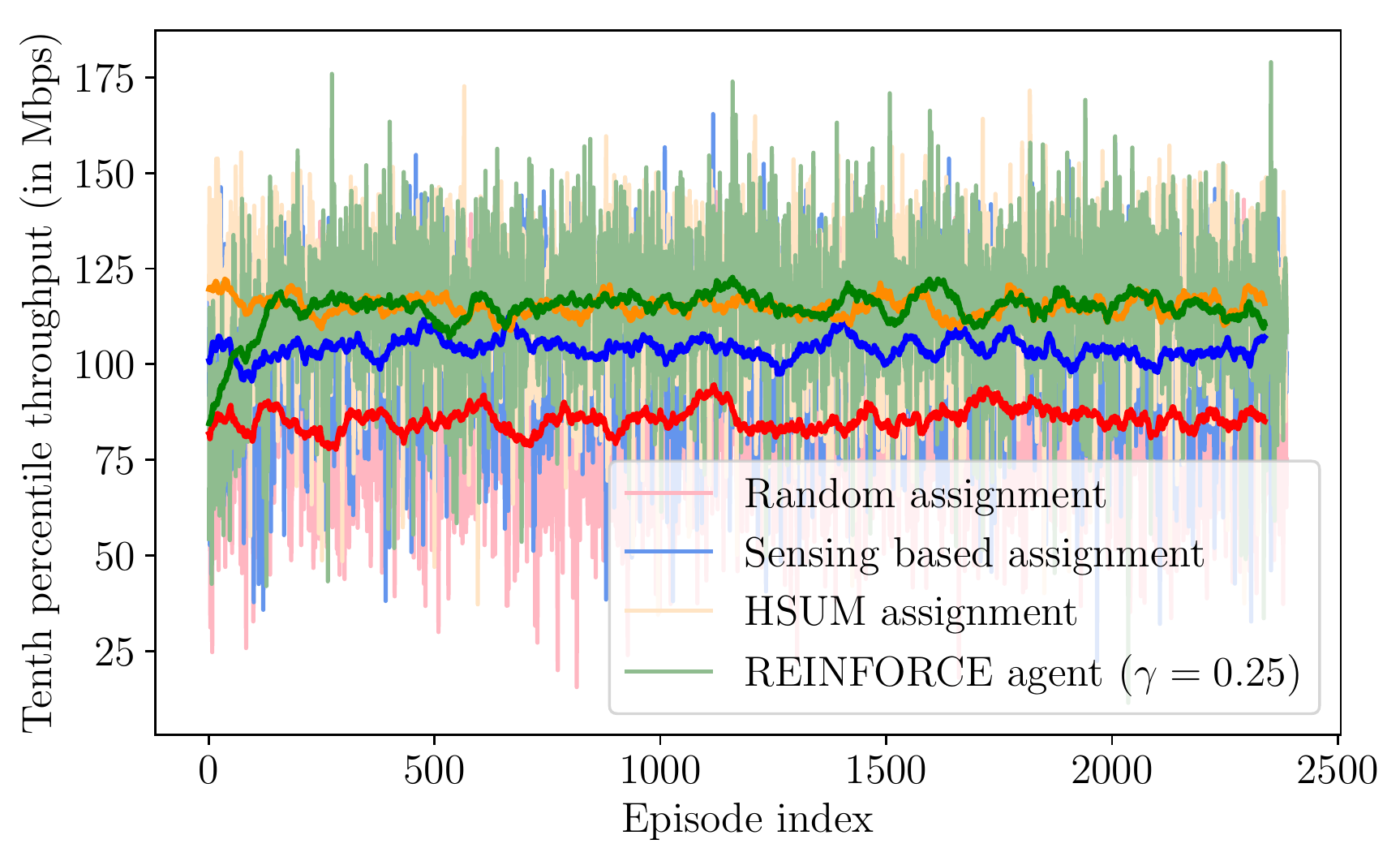}
		\caption{Results when one random external interferer was present} 
		\label{fig:P2_one_interferer} 
	\end{subfigure} \hfill
	\begin{subfigure}[t]{0.49\linewidth}
		\centering
		\includegraphics[width=\linewidth]{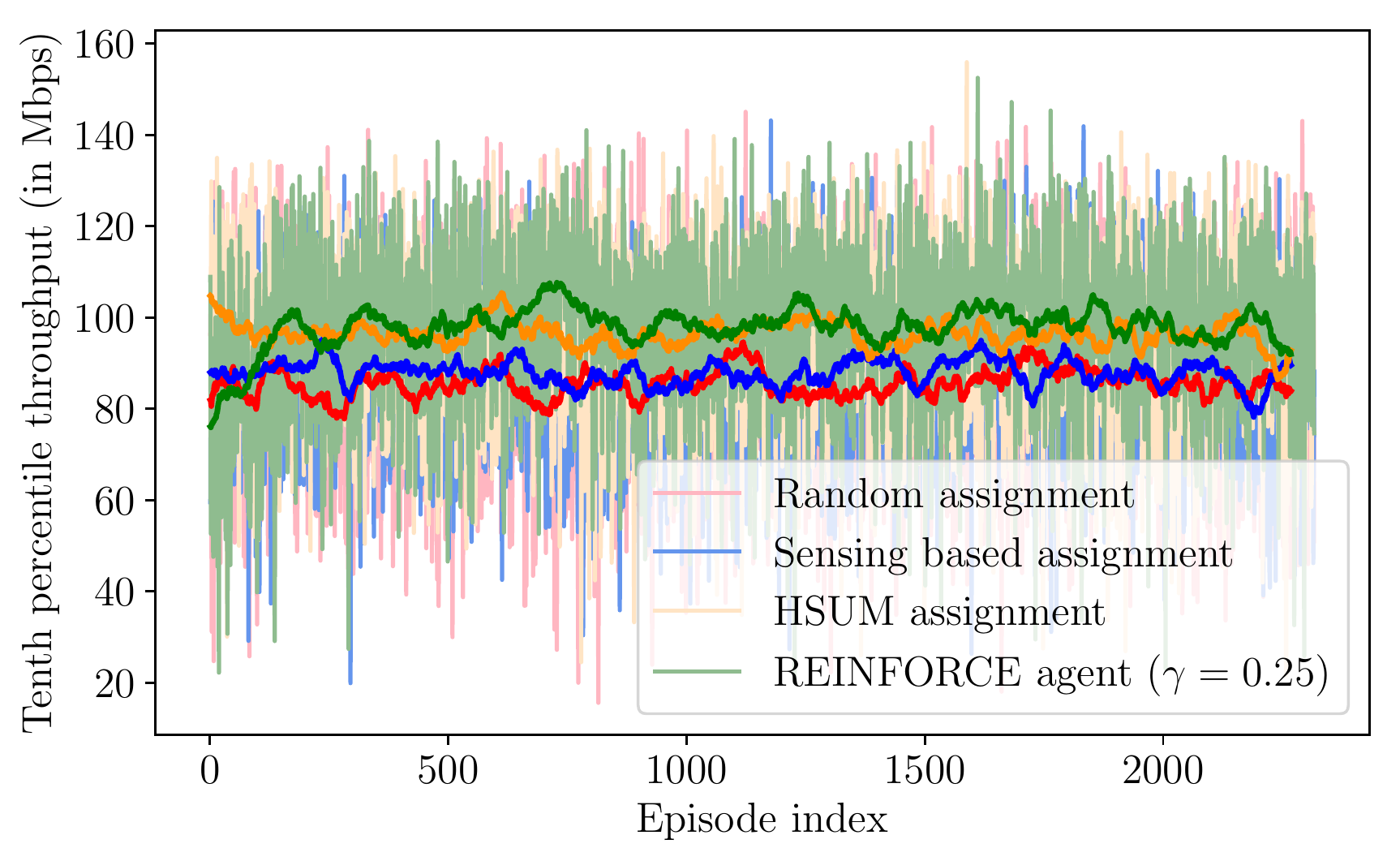}
		\caption{Results when three random external interferers were present} 
		\label{fig:P2_three_interferers} 
	\end{subfigure} 
	\caption{Throughput of the tenth percentile of users obtained using different channel assignment schemes in the presence of random external Wi-Fi interference}
	\label{fig:P2_results}
\end{figure*}
\subsection{Vanilla Channel Assignment}
\label{ssec:P1_results}
\noindent
\textbf{State} $s_t$: Group-channel assignments at step $t$ -- a $16 \times 1$ vector with element at index $i$ indicating the channel assigned to group $i$ \\
\textbf{Initial state of each episode} $s_0$: Channel red assigned to all $16$ groups \\
\textbf{Action} $a_t$: Change the channel assignment of one group \\
\textbf{Performance metric} $x_t$: Throughput of the thirtieth percentile of all users \\
\textbf{Reward} $r_t = x_t - x_{t-1}$ \\
\textbf{Agent}: On-policy REINFORCE (details of the implementation provided in Table~\ref{tab:specifics_of_REINFORCE_agent}) \\
First, consider the case of vanilla channel assignment in D-MIMO Wi-Fi networks (described in Section~\ref{ssec:P1}). The learning agent used policy gradients REINFORCE (Section~\ref{sssec:REINFORCE}) for training. To compare the results obtained from using the REINFORCE agent, the following channel assignment strategies were also considered : a) \emph{random assignment} in which D-MIMO groups were assigned channels randomly in every episode, and b) \emph{heuristic-based assignment} in which groups were assigned channels as shown in Fig.~\ref{fig:groups_different_channels}. Each episode began with the initial state $s_0$ (the worst-case channel assignment), an independent realization of the physical channels between $\RH$s and users, and an independent uniform distribution of users in the network space. 

Fig.~\ref{fig:P1_gamma_10} shows the throughput of the thirtieth percentile of all users observed over several episodes when using a DRL agent with discount factor $\gamma = 0.10$. The bold lines in Fig.~\ref{fig:P1_gamma_10} plot the moving averages over intervals of $50$ episodes and the lighter lines show the per-episode throughput. Notice how the DRL agent was able to attain a throughput value similar to when using the heuristic-based channel assignment within a few hundred episodes, even though each episode began with the worst case channel assignment strategy. Also observe the consistency of the DRL throughput in Fig.~\ref{fig:P1_gamma_10} after it has reached the peak, indicating that the variance of performance between episodes is minimal. To study the effect of discount factor, the agent was next used with discount factor $\gamma = 0.50$. Note from Fig.~\ref{fig:P1_gamma_50} that the agent was yet again successful in determining the best channel assignment for D-MIMO groups and the difference in discount factors determined the aggressiveness with which the agent reached the best channel assignment; the agent with $\gamma = 0.50$ took longer to reach the best assignment compared to the agent with $\gamma = 0.10$. Fig.~\ref{fig:P1_different_gamma} compares the performance of REINFORCE agents with different discount factors in determining the best channel assignment scheme and it is evident that agents with lower discount factors have better convergence compared to agents with higher values. This observation is in agreement with the findings reported in \cite{knox2012reinforcement} for episodic tasks.

\subsection{Channel Assignment with External Wi-Fi Interference}
\label{ssec:P2_results}
Next, we considered channel assignment in a D-MIMO Wi-Fi network but in the presence of external Wi-Fi interference in its vicinity (described in Section~\ref{ssec:P2}). The external interferers could be located within \SI{15}{m} (in x and y directions) of the D-MIMO network (as shown in Fig.~\ref{fig:dmimo_with_external_interference}). Note that \textbf{\emph{the location of the interferers, and the channels assigned to the interferers were different in each episode}} (channels assigned to different interferers were different as well). This may be interpreted as the network environment changing after each learning episode.

The definitions of state and action for this scenario were the same as in problem \ref{ssec:P1_results} and the performance metric in this case was the throughput of the tenth percentile of all users. Each episode began with the channel assignment determined by the DRL agent in scenario \ref{ssec:P1_results} (that is, when there was no external interference present). That is, if $s^*$ denotes the D-MIMO group channel assignment determined by the agent in Fig.~\ref{fig:P1_gamma_10}, then for each episode in the `with interference' scenario, the initial state of the environment for each episode was set to $s_0=s^*$. To compare the performance obtained using the DRL agent, two different channel assignment strategies were implemented: a) \emph{sensing based assignment} in which D-MIMO groups assigned channels to themselves (in a distributed manner) based on the energy sensed in each channel (note that all groups assign channels to themselves synchronously), and b) \emph{HSUM based assignment} \cite{mishra2005weighted} (after inducing necessary changes to the algorithm work in a distributed MIMO setting). 

The DRL learning agent used the REINFORCE algorithm with a discount factor $\gamma = 0.25$. Figs.~\ref{fig:P2_one_interferer} and \ref{fig:P2_three_interferers} describe the results when one and three random external interferers are present respectively. It can be observed that the DRL agent was able to update the channel assignments of the D-MIMO groups in response to the presence of external interferers. The DRL agent was able to achieve similar (in fact, better for most of the episodes) throughput numbers compared to when HSUM was used. These results demonstrate the success of the DRL agent in identifying the fact that groups near an interferers should be assigned a channel different from the interferers. Observe that even though each episode changed the location and channels of the interferers, the agent was still able to converge to the best policy within a few hundred episodes. 

\begin{figure*}[t] 
	\begin{subfigure}[t]{0.49\linewidth} 
		\centering
		\includegraphics[width=\linewidth]{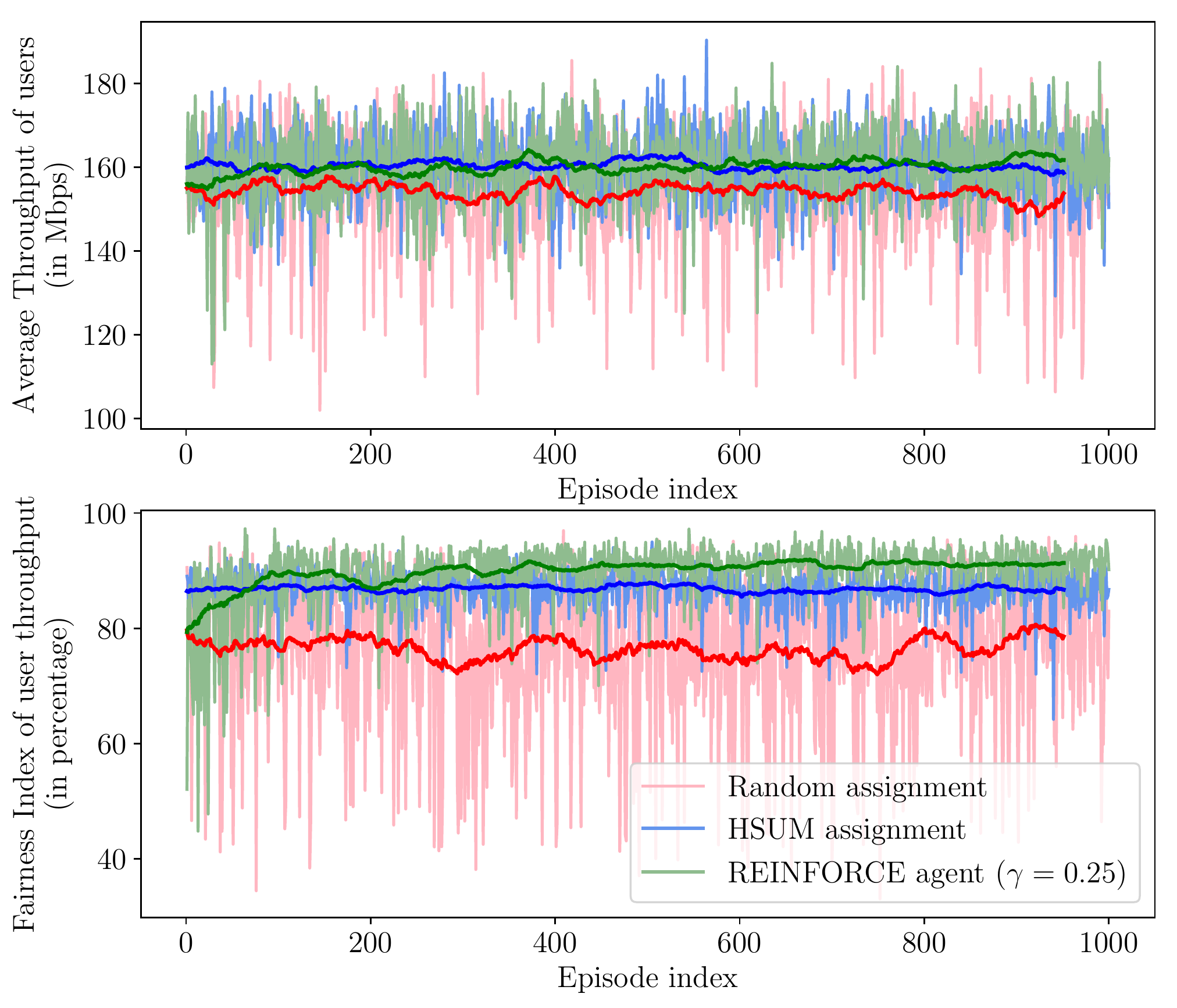}
		\caption{Results with three random external interferers} 
		\label{fig:P3_three_interferers} 
	\end{subfigure} \hfill
	\begin{subfigure}[t]{0.49\linewidth}
		\centering
		\includegraphics[width=\linewidth]{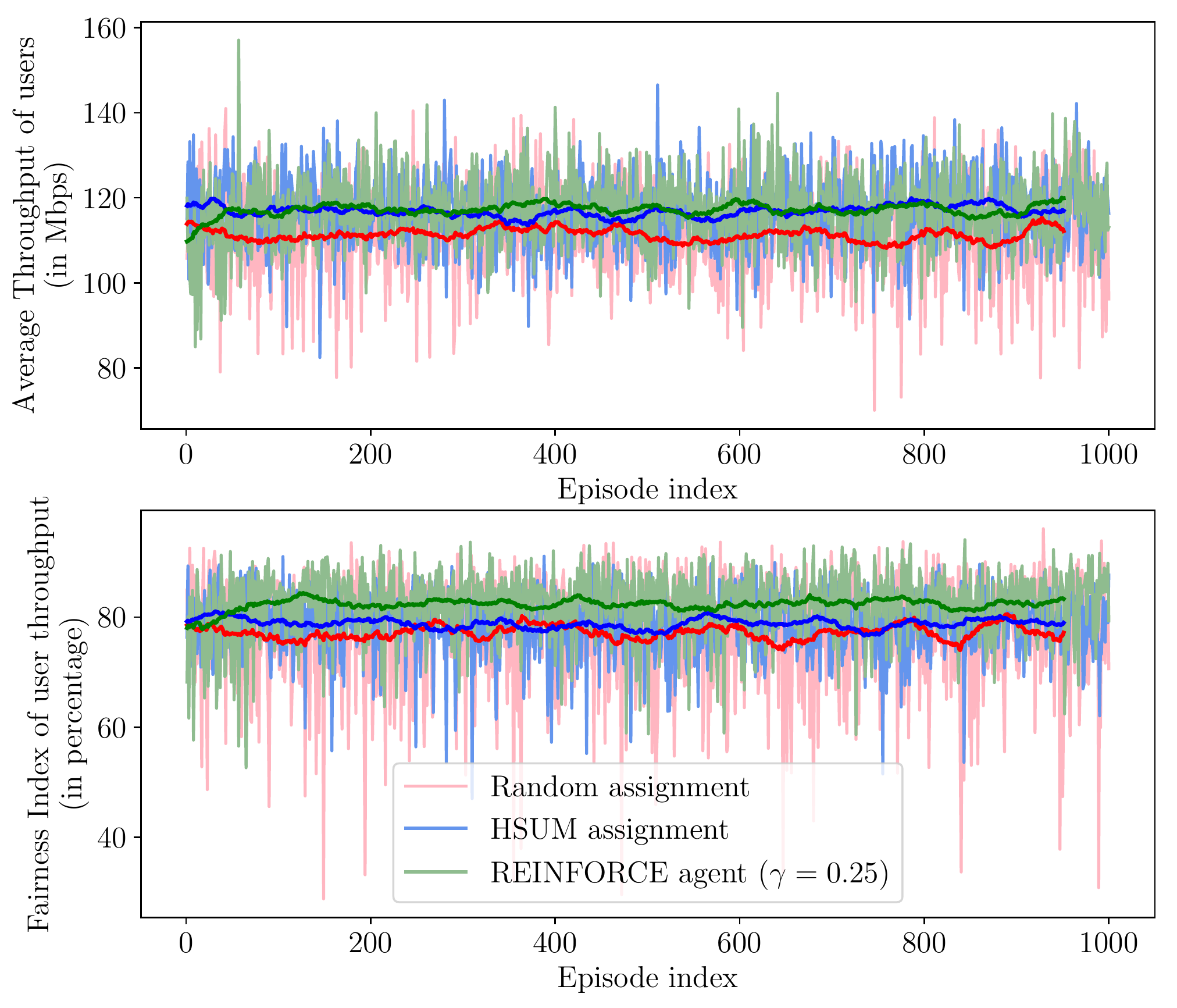} 
		\caption{Results with eleven random external interferers} 
		\label{fig:P3_eleven_interferers} 
	\end{subfigure} 
	\caption{Average throughput of users and Jain's fairness index of throughput among users in the presence of random external Wi-Fi interference}
	\label{fig:P3_results}
\end{figure*}

\subsection{Multiple Objectives}
\label{ssec:P3_results}
\noindent
\textbf{State} $s_t$ and \textbf{Action} $a_t$: Same as problems in Sections~\ref{ssec:P1_results} and \ref{ssec:P2_results}  \\
\textbf{Initial state of each episode} $s_0$: State $s^*$ as described in Section~\ref{ssec:P2_results} \\
\textbf{Performance metric} $x_t$ = Average throughput of users $\times$ Jain's fairness index of throughput among users \\
\textbf{Reward} $r_t$ = $x_t - x_{t-1}$ \\
\textbf{Agent}: On-policy REINFORCE agent; $\gamma=0.25$ \\
Consider the scenario of D-MIMO with external Wi-Fi interference but with two objectives -- maximize the average throughput of users (objective $O_1$) \textbf{\emph{and}} the fairness index of throughput among the users (objective $O_2$). Fairness, in this context, is the Jain's fairness index \cite{jain1999throughput}, defined as:
\begin{equation}
    \text{Fairness} = \frac{(\sum\limits_{i=1}^{n}x_i)^2}{n\cdot\sum\limits_{i=1}^{n}x_i^2}, 
\end{equation}
where $x_1,x_2,...,x_n$ represent the throughput numbers of users $1,2,...,n$. The objective of the DRL agent was to find a single policy which would meet both these objectives. The performance metric, in this case, was defined as the product of average throughput of users and the fairness index. The mapping of two performance measures to a single quantity is called \emph{scalarization}. If there exist $k$ objectives $O^1, O^2, ..., O^k$ with corresponding rewards $r^1, r^2, ..., r^k$, then scalarization condenses these $k$ rewards into a single metric $R = f(r^1, r^2, ..., r^k)$. There exist several methods for scalarization in literature -- linear scalarization (where the function $f$ is linear; usually a weighted sum of rewards), Chebyshev scalarization \cite{van2013scalarized} to name a few. However, in the current scenario of maximizing average user throughput and fairness index, these methods did not yield a good policy. This behavior may be ascribed to the disparity in the scale of the considered rewards -- fairness index was constrained to the range of $[0,1]$ while the throughput performance metric was not. 

Fig.~\ref{fig:P3_results} compares the performance of the DRL agent with HSUM-based channel assignment for two cases: three external interferers (Fig.~\ref{fig:P3_three_interferers}) and eleven external interferers (Fig.~\ref{fig:P3_eleven_interferers}). Note that, similar to the scenario considered in Section~\ref{ssec:P2_results}, \textbf{\emph{the location of interferers as well as the channels assigned to the interferers were changed after each episode.}} It can be observed that the DRL agent was able to achieve similar throughput performance as the HSUM-based assignment. However, the gains of using DRL are more evident in the trends of fairness index. The DRL agent was able to achieve higher throughput fairness among users while achieving a similar throughput performance as HSUM. Notice that the gains in fairness index are lower in Fig.~\ref{fig:P3_eleven_interferers} compared to Fig.~\ref{fig:P3_three_interferers}. This is understandable since there were more external Wi-Fi interferers in the vicinity of the D-MIMO network (in case of Fig.~\ref{fig:P3_eleven_interferers}) and hence the DRL agent had limited scope to update the channel assignments to improve the performance of the network. Even then, it was able to perform better than HSUM in terms of the achieved fairness index. The takeaway message from this section is that the agent was successful in (i) imbuing the D-MIMO network with resilience to dynamic external interference, and (ii) achieving a superior performance in simultaneously meeting the two objectives compared to HSUM.

\subsection{D-MIMO RH Grouping}
\label{ssec:P4_results}
\noindent
\textbf{State} $s_t$: $\RH$-group assignments (a $32\times1$ vector with element at index $i$ indicating the group to which $\RH$ $i$ belongs) concatenated with the user-$\RH$ assignments (a $50\times1$ vector with element at index $i$ representing the $\RH$ to which user $i$ is associated) at step $t$ \\
\textbf{Initial clustering of RHs in each episode}: According to the adjacent grouping strategy (see Fig.~\ref{fig:non_uniform_distribution_adjacent_grouping}) \\
\textbf{Action} $a_t$: Swap/exchange $\RH$ $a$ in group $b$ with $\RH$ $c$ in group $d$ \\
\textbf{Performance metric}: $x_t$ = Average throughput of users \\
\textbf{Reward}: $r_t$ = $x_t-x_{t-1}$ \\
\textbf{Agent}: Wolpertinger agent implementing DDPG for training (details of implementation in Table~\ref{tab:specifics_of_DDPG_agent}) \\
\begin{table}[t]
    \centering
    \caption{Specifics of the learning agent used in the Wolpertinger architecture in Section~\ref{ssec:P4_results}}
    \label{tab:specifics_of_DDPG_agent}
    \resizebox{\linewidth}{!}{
    \begin{tabular}{@{}ll@{}}
    \toprule
    Number of hidden layers & $2$ \\
    Number of input nodes & $82$ \\
    Number of output nodes & $1$ (Continuous proto action) \\
    \midrule  
    \multicolumn{2}{c}{\bfseries
  Actor Network Configuration} \\ \midrule 
    Number of hidden nodes & $256$ (first layer), $128$ (second layer) \\
    Learning rate ($\alpha_a$) & $0.003$ \\
    Target actor update parameter ($\tau_a$) & $0.001$ \\
    \midrule  
    \multicolumn{2}{c}{\bfseries
  Critic Network Configuration} \\ \midrule 
  Number of hidden nodes & $64$ (first layer), $32$ (second layer) \\
    Learning rate ($\alpha_c$) & $0.007$ \\
    Target critic update parameter ($\tau_c$) & $0.001$ \\
    \midrule
    Hidden node activation & Softplus \\ 
    Output node activation & Hyperbolic tangent (tanh) \\
    Optimizer & Adam \cite{kingma2014adam}\\
    Replay buffer size & $10,000$ samples \\
    Mini-batch size & $100$ \\
    \bottomrule
    \end{tabular}
}
\end{table}
As described in detail Section~\ref{ssec:P4}, this is a problem specific to D-MIMO-based Wi-Fi networks. Consider the D-MIMO network shown in Fig.~\ref{fig:DMIMO_non_uniform_distribution_of_users} with $32$ $\RH$s and eight groups (with four $\RH$s per group), with the $\RH$s clustered according to the adjacent grouping policy. Consider $50$ users non-uniformly distributed in the office space. The goal of the agent was to determine the best clustering policy in each episode based on the user distribution in that episode. \textbf{\emph{Note that the non-uniform distribution of users in the network was modified after each episode.}} This may be interpreted as the users being mobile and changing their locations at the end of every episode. The initial clustering method of $\RH$s at the beginning of each episode was the `adjacent grouping' arrangement as shown in Fig.~\ref{fig:non_uniform_distribution_adjacent_grouping}.

At each step, the agent performed an action which was to choose one pair of $\RH$s belonging to different groups and exchanged the $\RH$s between the groups. The reasoning behind defining action in such a manner was to maintain the number of $\RH$s per group as four always. This, in turn, was a deliberate decision to control the size of the action space. With such a definition of action, the number of actions from which the agent chose one, at each step, was $448$. Note that each episode began with $\RH$s clustered according to the adjacent grouping policy.

\underline{Note}: The users in the network were distributed such that some users clustered around a few (one/two) $\RH$s (as shown in Fig.~\ref{fig:DMIMO_non_uniform_distribution_of_users}). This was to purposely emulate the scenario of congregation of users in a conference/meeting room somewhere in the office space while the rest of the space was sparsely distributed with users. 

\begin{figure}[t]
    \centering
    \includegraphics[width=1\linewidth]{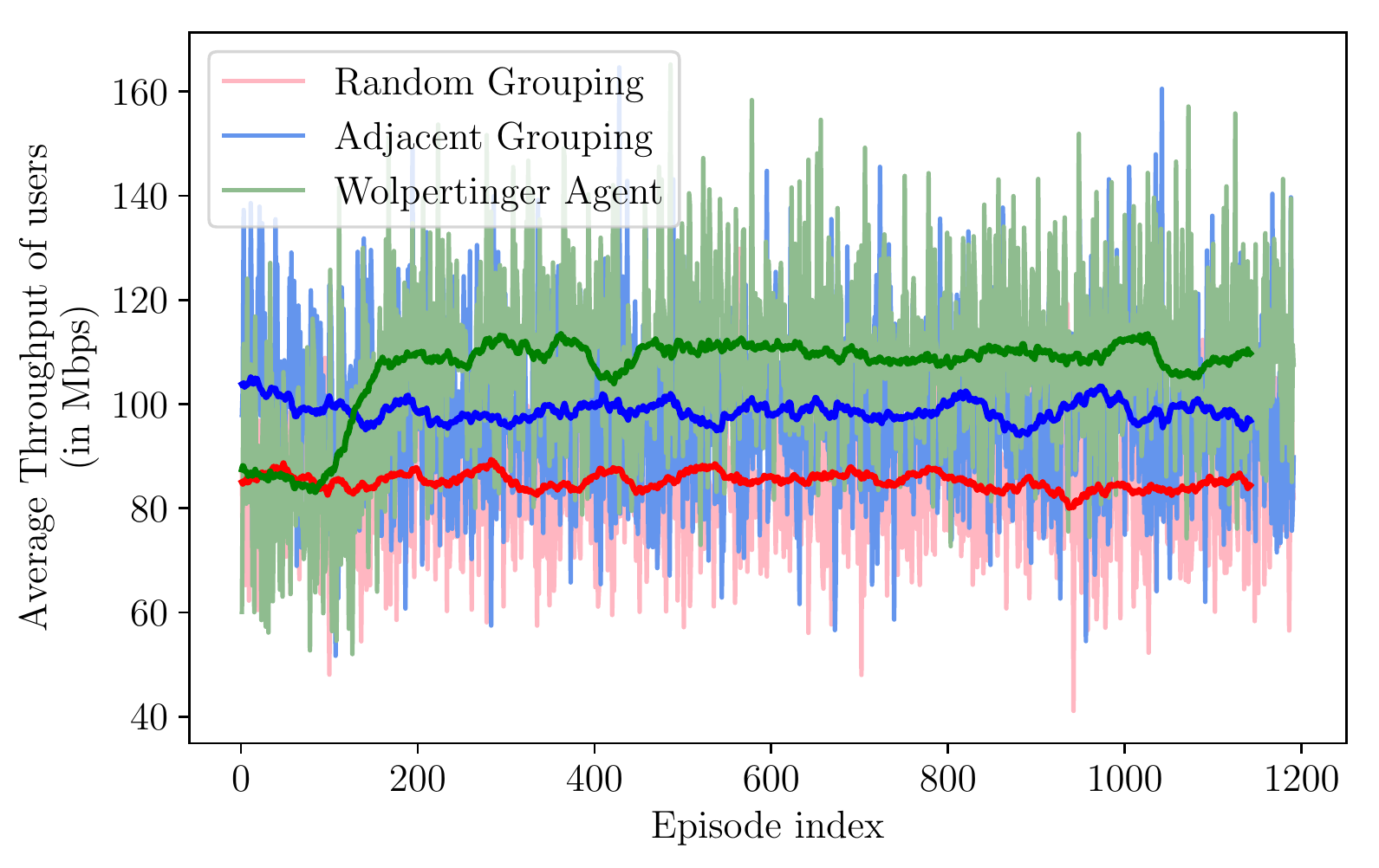}
    \caption{Average throughput of users when users were non-uniformly distributed in space}
    \label{fig:P4_results}
\end{figure}

The agent stored the observed trajectories/episodes in memory (called a \emph{replay buffer}) and randomly generated samples (called a \emph{mini-batch}) from this buffer which were used to perform its training. This idea is called \emph{experience replay} \cite{andrychowicz2017hindsight, lillicrap2015continuous, silver2016mastering} and it helped the agent achieve better convergence behavior and a higher sample efficiency. This is so because the training data are randomly sampled from the replay buffer and hence they act as uncorrelated data which help particularly in on-line training and non-linear function approximation.

Fig.~\ref{fig:P4_results} plots the average throughput performance of users when the Wolpertinger agent was used to update the $\RH$ grouping in response to the user distribution in each episode, along with results from two other clustering policies: i) when $\RH$s were randomly clustered in groups of four (referred to as \emph{random grouping}), and ii) when $\RH$s were clustered according to the \emph{adjacent grouping} strategy (as shown in Fig.~\ref{fig:non_uniform_distribution_adjacent_grouping}). Clearly, random clustering performed worse than the other two policies because it is blind to the distribution of users. It is also evident that the Wolpertinger agent was successful in reorganizing the clustering of $\RH$s to achieve a higher throughput performance compared to random clustering as well as the static adjacent grouping strategy; in fact, the Wolpertinger agent was able to attain an improvement of up to $20\%$ in the performance compared to the latter. This is encouraging because recent implementations of distributed MIMO \cite{hamed2016real,hamed2018chorus} have demonstrated the feasibility of achieving synchronization among $\RH$s belonging to a group over the air (without requiring a wired connection between them) and hence there is no logistical concern regarding clustering $\RH$s located far from each other. Also, observe that the performance of the agent was consistent across episodes even though the user distributions were changed in each episode. 



%% file: discussion.tex
\section{Discussion of Results and Future Research}
\label{sec:discussion}
Key takeaways from the results, and the lessons learned from implementing the DRL framework are summarized below: 
\begin{enumerate}
	\item DRL agents learned the optimal policies episodically. One learning episode consisted of a fixed number of actions. After the completion of one episode, the network environment was modified -- that is, the location of interferers, channels assigned to the interferers, and the distribution of users in the network space were different between different episodes. 
	\item DRL agents performed on-line training, that is, the agents learned with every observed episode (or with a random mini-batch of episodes in case of Section~\ref{ssec:P4_results}). Furthermore, the agents were fed \emph{simple state information} which could be easily obtained by the network administrator. This is desirable since traditional deep learning methods in the context of wireless networking rely heavily on collecting large amounts of data prior to training, the collection of which may be prohibitively expensive. Also, in the traditional deep learning case, there is a risk of developing a model which may over-fit to the training data if the amount of data collected is not sufficiently large and hence the model may not generalize. 
	\item DRL agents employed in this work had fairly simple configurations in terms of the number of hidden layers, number of hidden nodes, and the implemented training algorithm. 
	\item Using a single hidden layer in the implementation of the learning agent resulted in the following benefits (compared to agents with more hidden layers): (i) lower computational complexity, (ii) lower variance in performance between episodes, (iii) lower storage requirements (to store the parameters of the neural network), and (iv) faster training time. All of these aspects are attractive in the context of mobile networks as it reduces computational, memory, and energy requirements. Although using an agent with more hidden layers may improve training accuracy, the aforementioned advantages make a stronger case for using agents with a single hidden layer. 
    \item Maintaining a replay buffer with previously observed experiences and training the Wolpertinger agent based on a mini-batch randomly sampled from the buffer helped the agent achieve better convergence behavior and sample efficiency in Section~\ref{ssec:P4_results}. However, this comes at a price -- memory/storage overhead to store the replay buffer. It is encouraging to see the success of DRL agents in the channel assignment problems (in Sections~\ref{ssec:P1_results}, \ref{ssec:P2_results}, and \ref{ssec:P3_results}) without needing a replay buffer.
\end{enumerate}

The training results provided in Section~\ref{sec:results} serve as an impetus to continue studying the use of DRL to solve more complex and non-trivial problems in the context of wireless networks. Further research directions may be along the following lines:
\begin{itemize}
\item \emph{D-MIMO RH grouping}: While addressing the problem of $\RH$ grouping in D-MIMO networks in Section~\ref{ssec:P4_results}, a restriction on the number of $\RH$s per group was imposed. It will be interesting to study the performance of the DRL agent if such restrictions are relaxed, that is, the agent is free to cluster the $\RH$s in an unconstrained manner. However, this may lead to a longer learning time before convergence to the best policy since the agent will have to work with a very large action space. Additionally, channel assignment and $\RH$ grouping may be combined together, that is, a DRL agent may update the clustering of $\RH$s as well as the group-channel assignment in response to the distribution of users in the network.
\item \emph{Meeting multiple conflicting objectives}: The multiple objective problem in Section~\ref{ssec:P3} considered two objectives which went hand-in-hand with each other. However, there may exist simultaneous network objectives which need not be symbiotic, that is, improvements in one objective may lead to deterioration of others. In such cases, the multi-objective reinforcement learning framework becomes more complex with a single policy solution becoming highly improbable and the agent develops a convex convergence set of policies to meet different objectives of differing priority levels \cite{van2014multi}. 
\item \emph{Parameter settings for DRL}: The training of DRL agents involves extensive efforts of trial and error, especially in the setting of different hyper-parameters of the neural network, deciding the contents of the state information fed to the agent from the environment, and the formulation of rewards. Searching for the optimal configuration of the aforementioned parameters is analogous to looking for a needle in a haystack. Furthermore, these parameters have a high impact on the performance of the learning model. There is some recent work addressing this challenge by employing a progressive neural architecture search \cite{liu2018progressive} but this is a computationally expensive task.
\item \emph{Interpretability of learning algorithms}: Although it is evident from the training results that DRL agents were successful in performing better than heuristic solutions, we are still limited in our understanding of why the agents made certain decisions. This lack of interpretability has been widely regarded as a major reason impeding the pervasive application of DRL in the networking industry, albeit this is true in general with regard to its application in other domains as well. Active research has been undertaken to address this limitation and facilitate a better interpretability of learning algorithms \cite{chakraborty2017interpretability}.
\end{itemize}

%% file: conclusion.tex
\section{Conclusion}
\label{sec:conclusion}
This paper explored the potential of harnessing concepts from deep reinforcement learning (DRL) to enhance the performance of wireless networks. Specifically, this work focused on distributed multi-user MIMO (D-MIMO) Wi-Fi networks and addressed two major dynamic resource management problems in these networks: (i)  \emph{channel assignment of groups}, and (ii) \emph{clustering of radio heads to form groups}, in order to maximize user throughput performance. These problems are known to be NP-Hard for which only heuristic solutions exist in literature. A DRL framework was constructed to address the aforementioned problems and to enhance the performance of D-MIMO Wi-Fi networks when network conditions were dynamic. This paper considered practical dynamic network scenarios in which users were mobile and were distributed non-uniformly in space, and the network itself was subjected to random external Wi-Fi interference. This work considered DRL agents belonging to the policy iteration class, owing to the vastness of the action space of the considered scenarios. Through extensive simulations and on-line training based on D-MIMO Wi-Fi networks, this paper demonstrated that DRL agents successfully addressed the aforesaid problems as well as achieved an improvement of up to $20\%$ in user throughput performance compared to popular heuristic solutions. The DRL agents were also more effective, compared to heuristic solutions, in simultaneously meeting multiple network objectives (maximize throughput of users as well as fairness of throughput among them). 